\documentclass[10pt,final]{IEEEtran}
\usepackage{cite}

\ifCLASSINFOpdf
  \usepackage[pdftex]{graphicx}
\else
  \usepackage[dvips]{graphicx}
\fi
\usepackage{amssymb}
\usepackage[cmex10]{amsmath}
\usepackage{bm}
\usepackage{mathabx}
\usepackage{nccmath}
\usepackage{hyperref}                 
\usepackage{algorithm}
\usepackage{algpseudocode}

\usepackage{array}

\usepackage{mdwmath}
\usepackage{mdwtab}

\usepackage[caption=false,font=footnotesize]{subfig}
\usepackage{setspace}
\usepackage[usenames]{color}

\newcommand{\vect}[1]{\mathbf{#1}}
\usepackage{hhline}
\newtheorem{theorem}{Theorem}[section]
\newtheorem{definition}[theorem]{Definition}

\newtheorem{corollary}[theorem]{Corollary}
\newtheorem{proposition}[theorem]{Proposition}

\newenvironment{defn*}{\begin{definition}}{\end{definition}}

\newenvironment{proof}{\noindent{\bf Proof.}}{\hfill$\blacksquare$}
 \def\ci{{\perp\!\!\!\perp}} 

\newcommand{\thickbar}[1]{\mathbf{\bar{\text{$#1$}}}}

\begin{document}
 
\title{Latent Parameter Estimation in Fusion Networks Using Separable Likelihoods}
%

\author{Murat~\"{U}ney,~\IEEEmembership{Member,~IEEE,}
  Bernard Mulgrew,~\IEEEmembership{Fellow,~IEEE,}
  Daniel E. Clark,~\IEEEmembership{Member,~IEEE}
\thanks{This work was supported by the Engineering and Physical Sciences Research
Council (EPSRC) Grant number EP/K014277/1 and the MOD University Defence
Research Collaboration (UDRC) in Signal Processing.}
\thanks{Murat~\"{U}ney and Bernard Mulgrew are with the Institute for Digital Communications, School of Engineering, University of Edinburgh, EH9 3FB, Edinburgh, UK (e-mail: m.uney@ed.ac.uk, b.mulgrew@ed.ac.uk).}
\thanks{Daniel E. Clark is with D{\'e}partment CITI, Telecom-SudParis, 9, rue Charles Fourier
91011, EVRY Cedex, France (e-mail: daniel.clark@telecom-sudparis.eu).}
}

\markboth{IEEE Transactions on Signal and Information Processing Over Networks}%
{\"'{U}ney, Mulgrew and Clark: Latent Parameter Estimation in Fusion Networks Using Separable Likelihoods}

\maketitle
\begin{abstract}
Multi-sensor state space models underpin fusion applications in networks of sensors. Estimation of latent parameters in these models has the potential to provide highly desirable capabilities such as network self-calibration. Conventional solutions to the problem pose difficulties in scaling with the number of sensors due to the joint multi-sensor filtering involved when evaluating the parameter likelihood. In this article, we propose a separable pseudo-likelihood which is a more accurate approximation compared to a previously proposed alternative under typical operating conditions. In addition, we consider using separable likelihoods in the presence of many objects and ambiguity in associating measurements with objects that originated them. To this end, we use a state space model with a hypothesis based parameterisation, and, develop an empirical Bayesian perspective in order to evaluate separable likelihoods on this model using local filtering. Bayesian inference with this likelihood is carried out using belief propagation on the associated pairwise Markov random field. We specify a particle algorithm for latent parameter estimation in a linear Gaussian state space model and demonstrate its efficacy for network self-calibration using measurements from non-cooperative targets in comparison with alternatives.
\end{abstract}

\IEEEpeerreviewmaketitle
\section{Introduction}
\IEEEPARstart{A}{} wide range of sensing applications including wide area surveillance is underpinned by state space models which are capable of representing a variety of dynamic phenomena such as spatio-temporal (see, e.g.,~\cite{Sarkka2013}) processes. In fusion  (or, object tracking~\cite{Vo2015}) networks, multi-sensor versions of stochastic state space models, also known as hidden Markov models~\cite{Cappe2006}, are used to estimate object trajectories in a surveillance region.

These models, however, are often specified by some latent parameters~\cite{Cappe2007} some of which are unknown in practice and need to be estimated based on measurements from the state processes (or, objects). Examples of this problem setting in fusion networks include estimation of noise parameters~\cite{Singh2011a}, sensor biases~\cite{Lin2005,Ristic2013} and localisation/calibration of sensors in a GPS denying environment (e.g., in underwater sensing~\cite{Erol-Kantarci2011}) using point detections from non-cooperative targets~\cite{Kantas2012,Uney2014b}. Another example is the estimation of the orientations and positions of nodes in a camera network based on feature detections~\cite{Devarajan2008}.

Such problems fall in the domain of parameter estimation in state space models (see, e.g.,~\cite{Kantas2015} for a review). The parameter likelihood of the multi-sensor problem, however, does not scale well with the number of sensors which specifies the dimensionality of the unknown, or, the length of the measurement window that will be used for estimation. In the presence of multiple objects, the scalability issue is exacerbated by the measurement origin (or, data association) uncertainties that arise. Exact evaluation of the likelihood in this case has combinatorial complexity with the number of sensors~\cite{Chen2006}, and, in general multiple object models, it is intractable even for a single sensor~\cite{Jiang2015}. Estimation using a maximum likelihood (ML) or a Bayesian approach requires repeated evaluation of this likelihood (see, e.g.,~\cite{Kantas2015,Andrieu2010,Ala-Luhtala2016}) necessitating the use of efficient approximation strategies. 

Intractable or computationally prohibitive likelihoods have motivated a number of lines of work in the statistics literature including likelihood free methods, or, approximate Bayesian computation~\cite{Marin2012}, and, composite likelihood/pseudo-likelihood approaches~\cite{Varin2011}. Likelihood free methods can be used for sampling from the parameter posterior in state space models~\cite{Ehrlich2015} including those capable of modelling multiple objects~\cite{Yildirim2015}. The latter approach is based on developing surrogates to replace the original likelihood, e.g.,~block based approximations in maximum likelihood~\cite{Andrieu2005}. The pseudo-likelihood perspective has been useful in networked settings in which constraints on i) the availability of parts of data, and/or, ii) scalability in processing with the number of sources arise. Examples include surrogates built upon local functions for estimation of parametric probability measures (e.g., exponential family distributions) from distributedly stored high dimensional samples~\cite{Wiesel2012,Liu2012,Mizrahi2014}.

It is not straightforward to find such pseudo-likelihoods for parameter estimation in state space models, however, that can resolve these two issues that arise when there are multiple data sources (or, sensors). It is worthwhile to develop and analyse surrogates that provide scalability with the number of sources, and, are suitable to local computations (e.g., local filtering). In~\cite{Uney2014a}, we proposed a pseudo-likelihood which is a product of ``dual-term'' approximations replacing their intractable exact counterparts. These approximations are separable in that they can be evaluated using single sensor filtering. This feature underpins scalability with the number of sensors. In~\cite{Uney2016}, we have investigated the quality of the dual-term approximation, and, related it to the level of uncertainty in the prediction and estimation of the underlying state process.

In this work, we propose an alternative pseudo-likelihood which is provably a more accurate approximation for parameter estimation in multi-sensor state space models, under typical operating conditions. This approximation is also separable in that it is a scaled product of quadruple terms each of which can be found using single sensor filtering. In order to exploit this quad-term likelihood when there are multiple objects, extra attention should be paid to the handling of the data association uncertainties. We propose to use a hypothesis based parameterisation for the multi-object state space model as detailed in~\cite{Jiang2015}\&\cite{Yildirim2015a} in order to facilitate the use of the quad-term surrogate in this setting. In the parameterised model, we explicitly point out the combinatorial complexity of exact likelihood evaluation with the number of sensors. Then, we introduce an empirical Bayesian~\cite{Carlin2000} interpretation of local filtering that facilitates the use of separable likelihoods within this model. These modelling aspects detailing the use of separable likelihoods in hypothesis based multi-object models constitute the second contribution of this work. 

Separable likelihoods fit well in distributed fusion archictectures in which locally filtered distributions are transmitted in the network, as opposed to sensor measurements~\cite{Hall2013}. Moreover, they facilitate parameter estimation using a message passing computational structure which is desirable in networked problems. Specifically, the proposed likelihood surrogate together with independent parameter priors lead to a pairwise Markov random field (MRF) posterior model. The marginal distributions of this model approximates posterior marginals of the latent parameters to be estimated. We estimate these marginals iteratively using Belief Propagation (BP)~\cite{Yedida02} which consists of successive message passings among neighbouring nodes and updating of local marginals based on these messages. This computational structure lends itself to decentralised estimation, as well as scalable computation at fusion centre.

As an indication of the approximation quality, we consider the Kullback-Leibler divergence (KLD)~\cite{Cover1991} of the quad-term likelihood with respect to the actual pairwise likelihood and relate it to the uncertainties in predicting and estimating the underlying state using individual and joint sensor histories. We show that with more accurate local filters the approximation quality improves and the proposed quad-term separable likelihood has an improved error bound compared to the aforementioned dual-term approximation.

We provide a Monte Carlo algorithm for sensor self-calibration in this framework for linear Gaussian state space (LGSS) models. The algorithm is based on the nonparametric BP approach~\cite{Sudderth2010} and involves sampling from the updated marginals followed by quad-term likelihood evaluations in the message passing stage. As BP iterations converge to a fixed point, the empirical average of the samples from the marginals constitute (an approximate) minimum mean squared error (MMSE) estimate of the latent parameters. The edge potential are evaluated using the entire measurement history within a selected time period in an offline fashion which is a strategy similar to particle Markov chain Monte Carlo (MCMC) algorithms~\cite{Andrieu2010}. As such, we differ from~\cite{Uney2016} in which windowing of measurements are used 
for enabling online processing. 

Preliminary results of the proposed pseudo-likelihood can be found in~\cite{Uney2016a}. This article 
provides a complete account of our solution strategy in multiple object models and is structured as follows: Section~\ref{sec:problemdef} provides the probabilistic model and the problem statement. Then, we detail pairwise pseudo-likelihoods in parameterised multi-object models, and, relate this perspective to latent parameter estimation via inference over pairwise MRFs, in Section~\ref{sec:MRF}. The proposed quad-term node-wise separable likelihood approximation is detailed in Section~\ref{sec:seplhood}. Section~\ref{sec:quad4calibration} details the structural and computational properties of the quad-term approximation when the unknowns are respective quantities. Based on these results, we propose a distributed sensor localisation algorithm in linear Gaussian multi-object state space models in Section~\ref{sec:MonteCarloLBP}. The efficacy of this algorithm is demonstrated in comparison to the approach in~\cite{Uney2016}, in Section~\ref{sec:Example}. Finally, we conclude in~Section~\ref{sec:Discussion}.

\section{Problem Definition}
\label{sec:problemdef}
\subsection{Probabilistic model}
\label{sec:probmodel}
Let us consider a set of sensors ${\cal V}=\{1,\ldots,N\}$ networked over communication links listed by ${\cal E} \subset {\cal V}\times{\cal V}$. The graph ${\cal G} = ({\cal V},{\cal E})$ is undirected (i.e., the links are bi-directional), connected, and, might contain cycles. 

Next, let us consider a single object with state evolution modelled as a Markov process ${X}_k$ for time index $k \geq 1$. This process is specified by an initial state distribution and a transition density. The state space model with parameters $\theta$ is then specified as follows~\cite{Kantas2015}: The state value $x_k$ is a point in the state space $\mathfrak X$ and is generated by the chain
\begin{eqnarray}
 X_k | (X_{1:k-1} = x_{1:k-1}) &\sim &\pi( x_k | x_{k-1} ;\theta), \notag \\
 X_1 &\sim& \pi_b(x_{1};\theta),
 \label{eqn:statetransitionmodel}
\end{eqnarray}
where $.|.$ denotes conditioning. A measured value $z_k^i \in {\mathfrak Z}^i$ at sensor $i \in {\cal V}$ is generated independently in accordance with the likelihood model
\begin{eqnarray}
 Z_k^i | (X_{1:k} = x_{1:k}, Z^i_{1:k} = z_{1:k}^i) &\sim &g_i ( z_k^i | x_k  ;\theta ) 
 \label{eqn:measurementmodel}
\end{eqnarray}
where subscript $1:k$ indicates a vector concatenation over time.

In fusion scenarios, there are multiple such objects denoted by a multi-object state
\begin{equation}
 {\cal X}_k \triangleq \left[ X_{k,1},\dots, X_{k,M_k}  \right],
\end{equation}
that induce measurements according to the above state space model resulting with sensors collecting a multitude of measurements 
\begin{equation}
 {\cal Z}^i_k \triangleq \left[ Z_{k,1}^i,\dots, Z^i_{k,O^i_{k}}  \right],
\end{equation}
where $M_k$ is the number of objects and $O^i_{k}$ is the number of measurements collected at sensor $i$ at time $k$. Here, the origin of $Z_{k,j}^i$s are unknown, i.e., the {\it data association}s which encode a mapping from these measurement (random) variables to the elements of ${\cal X}_k$ (and, equivalently to the previously collected measurements from the same objects) are not known~\cite{Vo2015}.

In the general multi-object tracking model, $M_k$ and $O^i_{k}$ are random variables with laws determined by probability models regarding how these objects appear in the surveillance region and disappear (which is often referred to as their birth and death, respectively), the law for the false alarms, etc. For the sake of simplicity and ease of presentation in the limited space especially when relating computational complexity to the number of sensors and objects in the following discussion, we assume that all of the objects that exist at time step $k=1$ remain in the scene for the time window considered and there are no missed detections and false alarms in sensor measurements which imply that $M_k=M$ and $O^i_{k}=M_k$, respectively, for some positive integer $M$ with probability one.

In this simplified ``closed world'' model the multi-object state transition is given by
\begin{equation}
 \pi(\vect{X}_k | \vect{X}_{k-1}; \theta ) = \prod_{m=1}^M \pi(x_{k,m}|x_{k-1,m};\theta). \label{eqn:simpmopred}
\end{equation}

The likelihood of the measurements collected by sensor $i$ is conditioned not only on the multi-object state $\vect{X}_k$, but, also on a (data association) hypothesis $\tau^i_{k}$ that encodes the association of measurements to the  objects within $\vect{X}_k$:
\begin{equation}
 l_i( \vect{Z}^i_k | \vect{X}_k,\tau_k^{i}; \theta ) = \prod_{ o =1 }^M g_i( z^i_{k,o}| x_{k,\tau^i_k(o)} ;\theta),
 \label{eqn:simpmolhood}
\end{equation}
and, the prior on $\tau_k^{i}$ assigns equal probability to all $M!$ permutations of $[1,\ldots,M]$ that $\tau_k^{i}$ can take, i.e.,
\begin{equation}
 p(\tau_k^i) =\frac{1}{M!}.
 \label{eqn:equalikelyprior}
\end{equation}

\subsection{Statement of the problem}
\label{sec:statement}
We are interested in estimating $\theta \in {\cal B}$ using the measurements collected across the network by sensors $i \in {\cal V}$ for a time window of length $t$. The parameter likelihood of the problem quantifies how well these measurements fit into the state space model with the selected value of the parameter, and, is evaluated via multi-sensor filtering~\cite[Sec.IV]{Cappe2007}:
\begin{multline}
l\left({\vect{Z}^1_{1:t},\ldots,\vect{Z}^N_{1:t}|\theta}\right)=\\
\prod\limits_{k=1}^{t}p\left(\vect{Z}^1_{k},\ldots,{\vect{Z}^N_{k}|\vect{Z}^1_{1:k-1},\ldots,\vect{Z}^N_{1:k-1},\theta}\right), 
\label{eqn:centlhood}
\end{multline}
where the time updates on the right hand side are given by
\begin{multline}
p\left(\vect{Z}^1_{k},\ldots,{\vect{Z}^N_{k}|\vect{Z}^1_{1:k-1},\ldots,\vect{Z}^N_{1:k-1},\theta}\right)=
\\
\sum_{\tau^1_k}\cdots\sum_{\tau^N_k} p(\tau^1_k)\times \ldots \times p(\tau^N_k)\\
\times
\int_{ {\mathfrak{X}}^{M} } l({\vect{Z}^1_{k},\ldots,\vect{Z}^N_{k}} | \vect{X}_k,  \tau^1_k,\ldots,\tau^N_k ; \theta) 
\\
\times p(\vect{X}_k | \vect{Z}^1_{1:k-1},\ldots,\vect{Z}^N_{1:k-1};\theta ) \mathrm{d}{\vect{X}_k},
\label{eqn:factcent}
\end{multline}
and, the multi-sensor likelihood inside the integration factorises as
\begin{equation}
l({\vect{Z}^1_{k},\ldots,\vect{Z}^N_{k}} | \vect{X}_k, \tau^1_k,\ldots,\tau^N_k ; \theta)  = \prod_{i \in {\cal V}} l_i ( \vect{Z}_k^i | \vect{X}_k, \tau^i_k  ;\theta ).
\label{eqn:msensorlhood}
\end{equation}
where the terms in the product are given by~\eqref{eqn:simpmolhood}.

Here, \eqref{eqn:centlhood} follows from the chain rule of probabilities. The term in \eqref{eqn:factcent} is the contribution at time step $k$ which updates the likelihood of the previous time step and is found using the Markov property that the sensor measurements are mutually independent of the measurement histories, conditioned on the current state for any value of $\theta$. Let us denote this relation by ${{\cal{Z}}^j_{k} \ci {\cal{Z}}^j_{1:k-1}| {\cal{X}}_{k}, {\theta}}$ for $i \in {\cal V}$ (see, e.g.,~\cite{Wainwright2008}, for this notation). \eqref{eqn:msensorlhood} follows from that the measurements of different sensors are mutually independent, i.e.,  ${{\cal{Z}}^i_{k} \ci {\cal{Z}}^j_{k}| {\cal{X}}_{k}, {\theta}}$ for~${(i,j)\in {\cal V}\times {\cal V}}$.

This likelihood can be used in a MMSE estimator of $\theta \in {\cal B}$, in principle, for a random variable $\Theta$ associated with a prior density $p(\theta)$. This estimate is given by the expected value of the posterior distribution
\begin{eqnarray}
 p(\theta|\vect{Z}^1_{1:t},\ldots,\vect{Z}^N_{1:t})\,\, &\propto& \,\,l(\vect{Z}^1_{1:t},\ldots,\vect{Z}^N_{1:t}|\theta)\,p(\theta),
 \label{eqn:posthetaglobal}
 \\
 \hat \theta\,\, &=& \,\, \int_{\cal B} \theta \, p(\theta|\vect{Z}^1_{1:t},\ldots,\vect{Z}^N_{1:t})  \,\mathrm{d}\theta.
 \label{eqn:mmsetheta}
\end{eqnarray}
The MMSE estimate can be computed by generating $L$ samples from the posterior distribution in \eqref{eqn:posthetaglobal} using, for example, MCMC methods~\cite{Andrieu2010} and using these samples to find a Monte Carlo estimate of the integral in~\eqref{eqn:mmsetheta}. In both this approach and maximum likelihood (ML) solutions aiming to maximise~\eqref{eqn:centlhood} with iterative optimisation, repeated evaluations of the likelihood are required.

The evaluation of this likelihood is intractable, however, not only because of the $(M!)^N$ summations in~\eqref{eqn:factcent}, but, also because of the complexity in finding the integrations involved. The integrands here are i) the multi-sensor likelihood in~\eqref{eqn:msensorlhood}, and, ii) the prediction density for $\vect{X}_{k}$ based on the network's entire measurement history up to time $k$. In other words, \eqref{eqn:factcent} is the scale factor for the posterior density of Bayesian recursions, or, the ``centralised'' filter given~by
\begin{eqnarray}
\mspace{0mu}
p({\vect{X}_{k},\tau^{1:N}_k|\vect{Z}^1_{1:k},\ldots,\vect{Z}^N_{1:k}; \theta})= 
\mspace{200mu}
\nonumber \\
 \frac{l({\vect{Z}^1_{k},\ldots,\vect{Z}^N_{k}} | \vect{X}_k, \tau^{1:N}_k; \theta)}{{p\left({\vect{Z}^1_{k},\ldots,\vect{Z}^N_{k}|\vect{Z}^1_{1:k-1},\ldots,\vect{Z}^N_{1:k-1},\theta } \right)}} \times p(\tau^{1:N}_k)
 \mspace{0mu}
\nonumber \\
\times  p({\vect{X}_{k}|\vect{Z}^1_{1:k-1},\ldots ,\vect{Z}^j_{1:k-1},\theta}),
\mspace{50mu}
\label{eqn:update}
 \\
 \mspace{0mu}
p({\vect{X}_{k}|\vect{Z}^1_{1:k-1},\ldots,\vect{Z}^N_{1:k-1},\theta}) = 
\mspace{200mu} 
\notag \\
\sum_{\tau^1_{k-1}}\cdots\sum_{\tau^N_{k-1}} \int_{\mathfrak{X}^M} \pi(\vect{X}_k|\vect{X}_{k-1},\theta) 
\mspace{100mu}
\nonumber \\
\times p(\vect{X}_{k-1},\tau_{k-1}^{1:N}|\vect{Z}^1_{1:k-1},\ldots,\vect{Z}^N_{1:k-1},\theta) \mathrm{d}{\vect{X}}_{k-1},
\mspace{50mu}
\label{eqn:pred}
\end{eqnarray}
where we denote by $\tau_k^{1:N}$ the concatenation of $\tau_k^i$s and it has $(M!)^N$ different configurations. Here, both the prediction~\eqref{eqn:pred} and update~\eqref{eqn:update} are ${\cal O}((M!)^N)$.

In order to address these challenges, multi-object filtering (or, tracking) algorithms often employ two approximations: First, they aim to find the most probable data association  hypothesis in~\eqref{eqn:update} denoted by ${\bm {\thickbar{\tau}}}^{1:N}_k$, instead of both evaluating this expression for all possible associations and storing them. The benefits of doing so are that i)~one can generate tracks (or, object trajectories) as simply marginals of $p(\vect{X}_k,\tau^{1:N}_k = {\bm {\thickbar{\tau}}}^{1:N}_k | \,.\,)$ for $k=1,...,t$, and, ii)~evaluations of the integral in~\eqref{eqn:pred} in the next time step can be restricted to this value of the hypothesis variable. Equivalently, the posterior distribution in~\eqref{eqn:update} is factorised as
\begin{eqnarray}
&&\mspace{-30mu}p({\vect{X}_{k},\tau^{1:N}_k|\vect{Z}^1_{1:k},\ldots,\vect{Z}^N_{1:k}; \theta}) \nonumber \\
&=& 
p({\vect{X}_{k}|\vect{Z}^1_{1:k},\ldots,\vect{Z}^N_{1:k}, \theta,\tau^{1:N}_k})p(\tau^{1:N}_k| \vect{Z}^1_{1:k},\ldots,\vect{Z}^N_{1:k}, \theta) \nonumber \\
&=& p({\vect{X}_{k}|\vect{Z}^1_{1:k},\ldots,\vect{Z}^N_{1:k}, \theta, \tau^{1:N}_k} )p(\tau^{1:N}_k| \vect{Z}^1_{k},\ldots,\vect{Z}^N_{k}, \theta)
\label{eqn:conditionedposterior}
\end{eqnarray}
where the association variables appear as model parameters in the first term and the second term is similar to a prior distribution on these models with the difference that it is conditioned on the current measurements. At time $k-1$, let us select this ``empirical prior'' as 
\begin{equation}
 p(\tau^{1:N}_{k-1}| \vect{Z}^1_{k-1},\ldots,\vect{Z}^N_{k-1}, \theta) \leftarrow \delta_{{\bm {\thickbar{\tau}}}^{1:N}_{k-1}}(\tau^{1:N}_{k-1})
 \label{eqn:empiricalprior}
\end{equation}
where $\delta$ is Kronecker's delta function and $\leftarrow$ denotes assignment.
The second approximation follows from the first one: Evaluation of the prediction stage in~\eqref{eqn:pred} reduces to evaluation of the Chapman-Kolmogorov equation for only the most likely value of the association parameters. This approach is often referred to as empirical Bayes~\cite{Carlin2000}, and is used to facilitate approximate solutions to otherwise intractable problems. 

A similar approximation can be used when evaluating the parameter posterior in~\eqref{eqn:posthetaglobal}. This leads to the following likelihood
\begin{multline}
l\left({\vect{Z}^1_{1:t},\ldots,\vect{Z}^N_{1:t}|\theta,{\tau_{1:t}^{1:N}}}\right)=\\
\prod\limits_{k=1}^{t}p_{\tau_k^{1:N}}\left(\vect{Z}^1_{k},\ldots,{\vect{Z}^N_{k}|\vect{Z}^1_{1:k-1},\ldots,\vect{Z}^N_{1:k-1},\theta}\right), 
\label{eqn:conditionalcentlhood}
\end{multline}
conditioned on ${\tau_{1:t}^{1:N}}$ where the factors are defined by
\begin{multline}
p_{\tau_k^{1:N}}\left(\vect{Z}^1_{k},\ldots,{\vect{Z}^N_{k}|\vect{Z}^1_{1:k-1},\ldots,\vect{Z}^N_{1:k-1},\theta}\right) \triangleq\\
\int_{ {\mathfrak{X}}^{M} } l({\vect{Z}^1_{k},\ldots,\vect{Z}^N_{k}} | \vect{X}_k,   \tau^{1:N}_k; \theta) 
\\
\times p(\vect{X}_k | \vect{Z}^1_{1:k-1},\ldots,\vect{Z}^N_{1:k-1};\theta ) \mathrm{d}{\vect{X}_k}.
\label{eqn:factcent_emp}
\end{multline}
This likelihood evaluated at $\tau_{1:k}^{1:N} = {\bm {\thickbar{\tau}}}^{1:N}_{1:k}$ replaces the one in~\eqref{eqn:centlhood} when the empirical (model) prior is selected as in \eqref{eqn:empiricalprior} (see Appendix~\ref{sec:empiricalBayesParameterLikelihood} for details). We will refer to~\eqref{eqn:conditionalcentlhood} as the empirical likelihood. Note that~\eqref{eqn:factcent_emp} is the integral term in~\eqref{eqn:factcent}. 

The empirical likelihood update term is computationally more convenient, however, alone it is not sufficient for scalability with the number of sensors $N$: Finding ${\bm {\thickbar{\tau}}}^{1:N}_k$ is equivalently an $N+1$-dimensional assignment problem which is NP hard even for $N=2$ sensors~\cite{Poore2006} which partly underlies the local filtering paradigm for multi-sensor processing and our interest in compatible solutions. 
 For a moment, let us consider the problem for a single object, i.e., for $M=1$. In this case $\tau^i_k$ for $i=1,\ldots,N$ have only one possible configuration (i.e., there is no data association uncertainty).
Because the dimensionality of $\theta$ is specified by $N$ and~\eqref{eqn:centlhood} will be evaluated for roughly $NL$ samples (when estimating \eqref{eqn:mmsetheta}~(see, e.g., \cite{Ristic2004})) each of which costing -- in the simplest linear Gaussian measurements case~\eqref{eqn:factcent}~\footnote{Specifically, for linear Gaussian measurements with no data association uncertainty, the marginal parameter likelihood involves computation of the innovation covariance for the so called group-sensor measurements in joint multi-sensor filtering.}-- at the least $O(N^2t)$, the computational cost will be cubic in the number of sensors which can easily become prohibitive for large $N$.

The networked setting has additional constraints to take into account: The sensors perform local filtering of their measurements and exchange filtered (track) distributions over $\cal G$ as opposed to transmitting their measurements~\cite{Hall2013}. As a result, the network-wide measurements are not available to evaluate the likelihood of the problem. Instead, local distributions we denote by $p(\vect{X}_k,\tau^j_k= \hat \tau^j_k | \vect{Z}^j_{1:k} )$ are made available to neighbouring nodes where $\hat \tau^j_k$ is an approximation to the most probable association configuration ${\bm {\thickbar{\tau}}}^{j}_k$ found locally, based on only the local sensor measurements at sensor $j$. There are computationally efficient algorithms for finding such solutions for the single sensor problem~(see, e.g., \cite{Poore2006}~and the references therein). Therefore, a viable solution needs to build upon these densities and local data associations $\hat \tau^i_k$ as opposed to joint multi-sensor filtering in the network.

The problem we address in this work is the design of scalable approximations to~\eqref{eqn:centlhood} for estimating $\theta$ in a networked setting based on local filtering results at the nodes. The proposed approach also addresses the aforementioned computational bottleneck at fusion centres in centralised multi-sensor architectures with a designated node receiving unfiltered sensor measurements.

It is also worth noting that the parameter vector $\theta \in {\cal B}$ can be used to represent a wide variety of parameters of the global model 
some of which can be intrinsic to sensors $i \in {\cal V}$ individually such as parameters pertaining to local noise models. We are particularly interested in a second class of parameters which have dependencies among sensors such as respective parameters such as sensor locations and similar ``calibration'' parameters. In the former setting, the estimation of local parameters decouple into independent estimation problems which can be solved using a suitable approach (see, e.g.,~\cite{Yildirim2015a,Kokkala201584,Schlangen2017}). 

In our setting, $\theta \triangleq \left[\theta_1,\ldots,\theta_N \right]$ where $\theta_i$ is associated with $i \in {\cal V}$ and its estimation does not decouple and depends on all measurements across the network  due to the dependencies of parameters, which, in turn, brings forward the multi-sensor aspects of the problem this work aims to address. Because local filtering is performed, on the other hand, local estimation of data association $\hat \tau^i_k$ is available independent of $\theta$, which we discuss in detail later in Section~\ref{sec:quad4calibration}.

\section{A Pseudo-likelihood and a pairwise MRF posterior for decentralised estimation}
\label{sec:MRF}

Pseudo-likelihoods are constructed from likelihood like functions which are computationally convenient and defined typically over smaller subsets of the data to overcome difficulties posed by intractable likelihoods over the entire set of data~(see, e.g., ~\cite{Varin2011}~and the references therein). Let us denote the network-wide data set by 
\begin{equation}
\vect{Z} \triangleq [ \vect{Z}^1_{1:t}, \ldots, \vect{Z}^N_{1:t} ].
\nonumber
\end{equation}

A fairly general form for a pseudo-likelihood is given by~\cite{Varin2011}
\begin{equation}
 \tilde l(\vect{Z} | \theta ) = \prod_{s \in S} \tilde l( \vect{Z}_{d_s} | \vect{Z}_{c_s}, \theta )^{\omega_s }
 \label{eqn:PseudoLikelihood}
\end{equation}
where $S$ is an index set, $\omega_s$ is a positive real number, and, $\vect{Z}_{d_s}$ and $\vect{Z}_{c_s}$ are mutually exlusive subsets of $\vect{Z}$ (for example, $\vect{Z}_{d_1}=\vect{Z}^i_2$ and $\vect{Z}_{c_1}=\vect{Z}^j_{1}$, etc.). These sets can be selected in various ways ensuring that the factors are computationally convenient functions, for example, marginals and/or conditional densities, and, estimates based on $\tilde l(\vect{Z}|\theta)$ are sensible. Note that~\eqref{eqn:centlhood} is also in this form, however, with difficult to evaluate factors.


Let us consider $\theta = [\theta_1,...,\theta_N ]$ and a pseudo-likelihood surrogate for the empirical likelihood in \eqref{eqn:conditionalcentlhood}:
\begin{eqnarray}
\mspace{-20mu} \tilde l_{\tau^{1:N}_{1:t}}(\vect{Z}|\theta) &= &\prod_{(i,j)\in {\cal E}} l_{\tau^{i,j}_{1:t}}( \vect{Z}^i,\vect{Z}^j|\theta_{i,j} ),  \label{eqn:pseudolhood} \\
 &=&\prod_{(i,j)\in {\cal E}} \prod_{k=1}^t p_{\tau^{i,j}_k}( \vect{Z}^i_k,\vect{Z}^j_k|\vect{Z}^i_{1:k-1},\vect{Z}^j_{k-1},\theta_{i,j} ).
 \label{eqn:pseudolhoodUpdate}
\end{eqnarray}
Here, ${\cal E}$ is essentially the set of sensor pairs whose likelihoods we would like to incorporate into the pseudo-likelihood, and, it is convenient to choose them as those that share a communication link, in a networked setting (Section~\ref{sec:probmodel}).

\begin{figure}[t!]
  \centerline{
   \includegraphics[height=28mm]{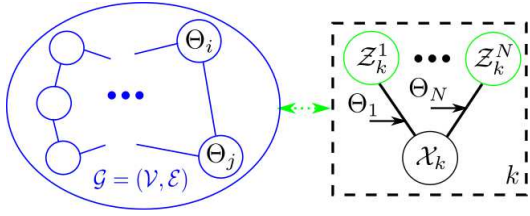}    
  }
  \caption[]{A multi-sensor state space - or, hidden Markov- model (black dashed box on the right representing a chain over $k$) and a Markov Random field model of the parameter posterior (the blue edges on the left).}
  \label{fig:distmodel}
\end{figure}

The pairwise structure above is beneficial to use with the MMSE estimator in \eqref{eqn:mmsetheta}. Note that the MMSE estimate is the concatenation of the expected values of posterior marginals, i.e., $p(\theta_i|\vect{Z})$ for $i=1,\ldots,N$. These distributions can be found using message passing algorithms over $\cal G$ when the surrogate \eqref{eqn:pseudolhood}~is used in~\eqref{eqn:posthetaglobal} together with independent but arbitrary {\it a priori} distributions selected for $\Theta_i$s. Specifically, the parameter posterior corresponding to such a selection of prioirs and the pseudo-likelihood~\eqref{eqn:pseudolhood} is a pairwise Markov random field over ${\cal G} = ({\cal V},{\cal E})$~\cite{Wainwright2008}:
\begin{eqnarray}
p(\theta | \vect{Z} ) &\propto &\prod_{i \in {\cal V}} \psi_i(\theta_i) \prod_{(i,j) \in {\cal E}} \psi_{ij}(\theta_i,\theta_j),
\label{eqn:approxpost}\\
\psi_i(\theta_i) &= &p_{0,i}(\theta_i), \nonumber \\
\psi_{ij}(\theta_i, \theta_j) &= & l_{\tau^{i,j}_{1:t}}( \vect{Z}^i, \vect{Z}^j | \theta_i,\theta_j ), \label{eqn:edgepot}
\end{eqnarray}
where the node potential functions (i.e., $\psi_i$s) are the selected priors (e.g., uniform distributions over  bounded sets $\theta_i$s take values from) 
and the edge potentials (i.e., $\psi_{ij}$s) are the pair-wise likelihoods for the pairs $(i,j)$s. This model is illustrated in \figurename~\ref{fig:distmodel}.

The pairwise MRF model in \eqref{eqn:approxpost} allows the computation of posterior marginal $p(\theta_i|{\vect{Z}})$ through iterative local message passings such as Belief Propagation (BP)~\cite{Yedida02}. In BP, the nodes maintain distributions over their local variables and update them based on messages from their neighbours which summarise the information neighbours have gained on these variables. This is described for all $i \in {\cal V}$ by
\begin{eqnarray}
m_{ji}( \theta_i )&=&\int \psi_{ij}(\theta_i, \theta_j)\, \psi_j(\theta_j) \prod\limits_{ i^\prime \in ne(j)\setminus i } m_{i^\prime j}(\theta_j)\,\mathrm{d}\theta_j,
 \label{eqn:bpmess} \\
\tilde p_i(\theta_i)&\propto& \psi_i(\theta_i)\prod_{j \in ne(i)} m_{ji}( \theta_i ).
 \label{eqn:bpstate} 
\end{eqnarray} 

In BP iterations, nodes simultaneously send messages to their neighbours using \eqref{eqn:bpmess} (often using constants as the previously received messages during the first step) and update their local ``belief'' using \eqref{eqn:bpstate}. If $\cal G$ contains no cycles (i.e.,  $\cal G$ is a tree), $\tilde p_i$s are guaranteed to converge to the marginals of \eqref{eqn:approxpost}, in a finite number of steps~\cite{Yedida02}. For the case in which $\cal G$ contains cycles, iterations of \eqref{eqn:bpmess}~and~\eqref{eqn:bpstate} are known as loopy BP (LBP). For the case, convergence does not have general guarantees, nevertheless LBP  has been been very successful in computing approximate marginals in a distributed fashion, in fusion, self-localisation and tracking problems in sensor networks~\cite{Ihler2005,Cetin2006,Uney2007}. In our problem setting, we assume that the models over spanning trees of a loopy $\cal G$ are consistent in that they lead to ``similar'' marginal parameter distributions, which suggests the existence of LBP fixed points~\cite{Wainwright2003} that will be converged when initial beliefs are selected reasonably~\cite{Yedidia2005}.

\section{Quad-term node-wise separable likelihoods}
\label{sec:seplhood}
The pseudo-likelihood introduced in Section~\ref{sec:MRF} leads to a parameter posterior that admits a pairwise MRF model. This is advantageous in providing a means for decentralised estimation through message passing algorithms in a network. The edge potentials~\eqref{eqn:edgepot} of this model, however, are i) conditioned jointly on two sensors' measurements simultaneous access to which is infeasible in a networked setting, and, ii) conditioned on association variables for two sensors and ideally should be evaluated at its most probable configuration  ${\bm {\thickbar{\tau}}}^{i,j}_{1},\ldots,{\bm {\thickbar{\tau}}}^{i,j}_{t}$ each of which is NP-hard to find, as explained in Section~\ref{sec:problemdef}.

In order to overcome these difficulties, we introduce an approximation which factorises into terms local to nodes, i.e., a node-wise separable approximation. 
Let us consider the ``centralised'' pairwise likelihood update term in~\eqref{eqn:pseudolhoodUpdate} given some configuration $\tau^{i,j}_k$ for $k=1,\ldots,t$, and, drop them from the subscript for the sake of simplicity in notation, as well as the $i,j$ subscript in $\theta$, in the following discussion. 
This term factorises in alternative ways as follows:
\begin{eqnarray}
 && \mspace{-20mu} p(\vect{Z}_k^i, \vect{Z}_k^j| \vect{Z}_{1:k-1}^i,\vect{Z}_{1:k-1}^j, \theta ) \notag \\
  &=& p(\vect{Z}_k^i| \vect{Z}_{1:k-1}^i,\vect{Z}_{1:k}^j, \theta ) p( \vect{Z}_k^j |\vect{Z}_{1:k-1}^i,\vect{Z}_{1:k-1}^j, \theta )
  \label{eqn:factzi} \\
  &=& p(\vect{Z}_k^j| \vect{Z}_{1:k}^i, \vect{Z}_{1:k-1}^j, \theta ) p( \vect{Z}_k^i |\vect{Z}_{1:k-1}^i,\vect{Z}_{1:k-1}^j, \theta )
  \label{eqn:factzj}\\
  &=& \left( p(\vect{Z}_k^i| \vect{Z}_{1:k-1}^i,\vect{Z}_{1:k}^j, \theta ) p( \vect{Z}_k^j |\vect{Z}_{1:k-1}^i,\vect{Z}_{1:k-1}^j, \theta ) \right)^{1/2} \notag \\ 
  && \!\!\!\times \left( p(\vect{Z}_k^j| \vect{Z}_{1:k}^i,\vect{Z}_{1:k-1}^j, \theta ) p( \vect{Z}_k^i |\vect{Z}_{1:k-1}^i,\vect{Z}_{1:k-1}^j, \theta ) \right)^{1/2}
  \label{eqn:geomean}
\end{eqnarray}

In the first and second lines above, the chain rule is used. The third equality can be found by taking the geometric mean of the first two expressions. All four factors in Eq.\eqref{eqn:geomean} are conditioned on the measurement histories of both sensors to which one cannot have simulatenous access in a networked setting. We would like to aviod this by leaving out the history of sensor $i$ (sensor $j$) in the first two (last two) terms of \eqref{eqn:geomean},~i.e.,
\begin{multline}
q(\vect{Z}_k^i, \vect{Z}_k^j| \vect{Z}_{1:k-1}^i,\vect{Z}_{1:k-1}^j, \theta ) \\
 \triangleq \frac{1}{\kappa_k(\theta )} \left( p(\vect{Z}_k^i| \vect{Z}_{1:k}^j, \theta ) p( \vect{Z}_k^j |\vect{Z}_{1:k-1}^j, \theta ) \right)^{1/2}  \\ 
 \times \left( p(\vect{Z}_k^j| \vect{Z}_{1:k}^i, \theta ) p( \vect{Z}_k^i |\vect{Z}_{1:k-1}^i, \theta ) \right)^{1/2}
 \label{eqn:quadapprox}
\end{multline}
\begin{multline}
\kappa_k(\theta )= \int \int \left( p( {\vect{Z}_k^i}' ,{\vect{Z}_k^j}'|\vect{Z}_{1:k-1}^j, \theta) \right. \\
\left. \times p( {\vect{Z}_k^i}' , {\vect{Z}_k^j}' |\vect{Z}_{1:k-1}^i, \theta ) \right)^{1/2} {\mathrm{d}{\vect{Z}_k^i}'\mathrm{d}{\vect{Z}_k^j}'}
\label{eqn:quadscale}
\end{multline}
where $\kappa_k(\theta )$ is the normalisation constant that guarantees $q$ to integrate to unity. Note that $\kappa_k$ is a function of the parameters~$\theta$.

The appeal of this quadruple term is that its factors depend on single sensor histories. As such, they require filtering of sensor histories of $i$ and $j$ individually enabling the evaluation of their product in a network. This point is discussed later in this section. 

\subsection{Approximation quality}
We consider the difference between the original centralised update term in \eqref{eqn:geomean} and the quad-term approximation introduced in~\eqref{eqn:quadapprox}. Because these terms are probability densities over sensor measurements, their ``divergence'' can be quantified using the KLD~\cite{Cover1991}:

\begin{proposition}
\label{prop:quadupperbound}
 The KLD between the centralised update and the node-wise separable approximation in \eqref{eqn:quadapprox} is bounded by the average of the mutual information (MI)~\cite{Cover1991} between the current measurement pair and a single sensor's history conditioned on the history of the other sensor, i.e.,
 \vspace{-4pt}
 \begin{multline}
 \mspace{-20mu}
D(p(\vect{Z}_k^i, \vect{Z}_k^j| \vect{Z}_{1:k-1}^i,\vect{Z}_{1:k-1}^j,\theta)||q(\vect{Z}_k^i, \vect{Z}_k^j| \vect{Z}_{1:k-1}^i,\vect{Z}_{1:k-1}^j,\theta) )  \\ 
\leq \frac{1}{2}I({\cal Z}_k^i, {\cal Z}_k^j; {\cal Z}_{1:k-1}^i | {\cal Z}_{1:k-1}^j,{\Theta} ) \\
+ \frac{1}{2}I({\cal Z}_k^i, {\cal Z}_k^j;{\cal Z}_{1:k-1}^j| {\cal Z}_{1:k-1}^i  ,{\Theta} ).
 \label{eqn:quadupperbound}
\end{multline}

\end{proposition}
The proof can be found in Appendix~\ref{sec:proofProp1} \footnote{Note that the results presented in this section are valid for any selection of ${\tau}^{i,j}_{1:t}$ as they relate random variables which are conditioned on the data association. The divergences and bounds, nevertheless, are more relevant for~${\tau}^{i,j}_{1:t} = {\bm {\thickbar{\tau}}}^{i,j}_{1:t}$.}. The upper bound in~\eqref{eqn:quadupperbound} measures the departure of the current pair of measurements, and, one of the sensor histories from a state of conditional independence when they are conditioned on the history of the other sensor. Note that these variables, when conditioned on ${\cal X}_k$, are conditionally independent, i.e.,  $({\cal Z}_k^i, {\cal  Z}_k^j) \ci {\cal  Z}^j_{1:k-1}|{\cal  X}_k,{\Theta}$ holds and consequently

\begin{equation}
 I({\cal Z}_k^i, {\cal  Z}_k^j; {\cal  Z}_{1:k-1}^i | {\cal  X}_{k},{\Theta} ) = I({\cal Z}_k^i, {\cal Z}_k^j; {\cal Z}_{1:k-1}^j | {\cal X}_{k},{\Theta}) = 0. \notag
\end{equation}
Similarly, the average MI term on the right hand side of \eqref{eqn:quadupperbound} is zero if $ ({\cal Z}_k^i, {\cal Z}_k^j) \ci {\cal Z}^i_{1:k-1}|{\cal Z}^j_{1:k-1} ,{\Theta}$ and $({\cal  Z}_k^i, {\cal Z}_k^j) \ci {\cal Z}^j_{1:k-1}|{\cal Z}^i_{1:k-1},{\Theta}$ hold simultaneously. This condition is satisfied, for example, in the case that either of the measurement histories ${\cal Z}^i_{1:k-1}$ and ${\cal Z}^j_{1:k-1}$ are sufficient statistics for ${\cal X}_k$ (i.e., it can be predicted by both sensors with probability one). This level of accuracy should not be expected as the transition density of state space models introduce some uncertainty. Therefore, it is instructive to relate the KLD in \eqref{eqn:quadupperbound} further to the uncertainty on ${\cal X}_k$ given the sensor histories:
\begin{corollary}
 \label{cor:quadupperbound}
 The KLD considered in Proposition~\ref{prop:quadupperbound} is upper bounded by the weighted sum of uncertainty reductions in the local target prediction and posterior distributions achieved when the other sensor's history is included jointly:
 \begin{multline}
\mspace{-20mu}
D(p(\vect{Z}_k^i, \vect{Z}_k^j| \vect{Z}_{1:k-1}^i,\vect{Z}_{1:k-1}^j,\theta)||q(\vect{Z}_k^i, \vect{Z}_k^j| \vect{Z}_{1:k-1}^i,\vect{Z}_{1:k-1}^j,\theta) )  \\ 
\leq  \frac{1}{2}\Bigg( \left( H({ \cal X}_k|{  \cal Z}^j_{1:k-1},\Theta )-H({ \cal  X}_k|{ \cal  Z}^j_{1:k-1},{ \cal  Z}^i_{1:k-1},\Theta ) \right)  \\
 \,\,+ \left( H({ \cal  X}_k|{ \cal  Z}^i_{k-1},\Theta ) - H({ \cal  X}_k|{ \cal  Z}^j_{1:k-1},{  \cal Z}^i_{1:k-1},\Theta ) \right) \Bigg)  \\
 \,\,\,\,
 + \frac{1}{2}\Bigg( \left( H({ \cal X}_k|{\cal  Z}^j_{1:k},\Theta )-H({\cal  X}_k|{\cal  Z}^j_{1:k},{\cal  Z}^i_{1:k-1},\Theta ) \right)  \\
 +\left( H({\cal  X}_k|{\cal  Z}^i_{1:k},\Theta )-H({\cal  X}_k|{\cal  Z}^i_{1:k},{\cal  Z}^j_{1:k-1},\Theta ) \right)   \Bigg),
 \label{eqn:corolH} \\[-22pt] 
 \end{multline}
 where $H$ denotes the Shannon differential entropy~\cite{Cover1991}.
\end{corollary}
The proof is provided in Appendix~\ref{sec:proofCorollary1}. Corollary~\ref{cor:quadupperbound} relates the approximation quality of the quad-term node-wise separable updates to the uncertainties in the target state prediction and posterior distributions when individual node histories and their combinations are considered. The difference terms on the RHS of~\eqref{eqn:corolH} quantify the difference in uncertainty between estimating the target state ${\cal X}_k$ using only the local measurements, and, also taking into account the other sensor's measurements. Overall, a better quality of approximation should be expected when the local filtering densities involved concentrate around a single point in the state space. 

\subsection{The quad-term pairwise likelihood}

The quad-term update in~\eqref{eqn:quadapprox} leads to a separable approximate likelihood given by
\begin{equation}
\tilde l\left({\vect{Z}^i_{1:t},\vect{Z}^j_{1:t}|\theta}\right) =\prod_{k=1}^{t} q(\vect{Z}^i_k,\vect{Z}^j_k|\vect{Z}^i_{1:k-1},\vect{Z}^j_{1:k-1},\theta )
\label{eqn:quadpotwnorm}
\end{equation}

We refer to this term as the quad-term separable likelihood as it can also be expressed as a (scaled) product of four factors each of which are the products of the four factors of~\eqref{eqn:quadapprox} over~$k$. Let us define
\begin{eqnarray}
 r^k_{ij}({\vect Z}_k^i,\theta) &\triangleq& p({\vect Z}_k^i| {\vect Z}_{1:k}^j, \theta ),
 \nonumber
 \\
 s^k_{j}({\vect Z}_k^j,\theta) &\triangleq &p({\vect Z}_k^j|{\vect Z}_{1:k-1}^j,\theta).
 \nonumber
\end{eqnarray}
Then, the quad-term update in~\eqref{eqn:quadpotwnorm} is given by
\begin{multline}
q(\vect{Z}^i_k,\vect{Z}^j_k|\vect{Z}^i_{1:k-1},\vect{Z}^j_{1:k-1},\theta ) = \\
\frac{1}{\kappa_k(\theta)} \left( r^k_{ij}({\vect Z}_k^i,\theta) s^k_{j}({\vect Z}_k^j,\theta)\right)^{1/2}\!\!\left( r^k_{ji}({\vect Z}_k^j,\theta) s^k_{i}({\vect Z}_k^i,\theta)\right)^{1/2},
\label{eqn:quadupdate2}
\end{multline}
where the normalisation factor is given in~\eqref{eqn:quadscale}, and, equivalently in terms of the four factors above as
\begin{multline}
 \kappa_k(\theta) = \int \int \left( r^k_{ij}( {{\vect Z}_k^i}',\theta) s^k_{j}({{\vect Z}_k^j}',\theta)\right)^{1/2} \\
 \times \left( r^k_{ji}({{\vect Z}_k^j}',\theta) s^k_{i}({{\vect Z}_k^i}',\theta)\right)^{1/2} {\mathrm{d}{{\vect Z}_k^i}'\mathrm{d}{{\vect Z}_k^j}}'.
\notag 
\end{multline}

\begin{corollary}
 The KLD between the parameter likelihood in \eqref{eqn:centlhood} and the node-wise separable approximation in \eqref{eqn:quadpotwnorm} is bounded by the terms on the right hand sides of \eqref{eqn:quadupperbound}~and~\eqref{eqn:corolH} summed over $k=1,\dots,t$ as
 \begin{multline}
  D\left( l\left({{\vect Z}^i_{1:t},{\vect Z}^j_{1:t}|\theta}\right) \vert\vert \tilde  l\left({{\vect Z}^i_{1:t},{\vect Z}^j_{1:t}|\theta}\right) \right) =   \sum_{k=1}^t \! D\left(p||q \right).
  \label{eqn:kldlhood}
 \end{multline}
\end{corollary}
\begin{proof}
Eq.~\eqref{eqn:kldlhood} can easily be found after expanding the KLD term explicitly and expressing the logarithm of products involved as sums over logarithms of the factors. Boundedness follows from non-negativity of KLDs and summing both sides of \eqref{eqn:quadupperbound}~and~\eqref{eqn:corolH} over $k=1,...,t$. 
\end{proof}

As a conclusion, when $t$ is not large -- e.g., on the order of tens which is typical in fusion applications-- the proposed approximation can be used for parameter estimation via local filtering. Sensors that are more accurate in inferring the underlying state process result with a smaller KLD in~\eqref{eqn:kldlhood}, which in turn leads to a more favourable estimation performance. 

One other approximation based on local filtering distributions was studied in~\cite{Uney2016} which has the following dual-term product form
\begin{multline}
 u( \vect{Z}_k^i, \vect{Z}_k^j| \vect{Z}_{1:k-1}^i,\vect{Z}_{1:k-1}^j,\theta ) \\ \triangleq p( \vect{Z}_k^i | \vect{Z}_{1:k-1}^j,\theta )p( \vect{Z}_k^j|  \vect{Z}_{1:k-1}^i,\theta ).
 \label{eqn:dualtermupdate}
\end{multline}

In Appendix~\ref{sec:kldcomp}, we shown that $D(p||q) < D(p||u)$, when sensors are equivalent. The entropy bound given in \eqref{eqn:corolH} is also smaller than that for the dual-term approximation. In other words, the quad-term approximation is more accurate compared to the dual-term approximation, under typical operating conditions. 

The scaling factor of the dual term approximation is unity regardless of $\theta$, on the other hand, admitting a significant amount of flexibility in the range of the distributions and likelihoods that can be accommodated in the state space model. For example, the dual-term pseudo-likelihood is used with random finite set variables (RFS) in~\cite{Uney2016,Uney2016b}, which, in a sense, have the association variables marginalised out making it possible to avoid  multi-dimensional assignment problems in the general multi-object tracking model.
For RFS distributions, however, it is not straightforward to compute the scaling factor in~\eqref{eqn:quadscale} for the quad-term. In this article, we consider a parametric model instead, which is effectively configured through association variables. 

\section{Quad-term likelihood for sensor calibration parameters}
\label{sec:quad4calibration}
The results presented so far are fairly general and do not depend on the nature of $\Theta$. When $\Theta$ represents respective parameters such as calibration parameters, there are certain simplifications of the expressions involved which provide computational benefits. In particular, parameters such as respective location and bearing angles relate the local coordinate frames of the sensors which collect measurements in their local frame. The local filtering distributions are hence over the space of state vectors in the local frame. A point $x_k \in \mathfrak X$ (Section~\ref{sec:probmodel}) is implicitly in the Earth coordinate frame (ECF), and, associated with its representation in the $j$th local frame $[x_k]_j$ through a coordinate transform $T$ with the following properties
\begin{eqnarray}
 [ x_k]_j &= & T(x_k;\theta_j ),
 \label{eqn:coortrans_tau} \\
 [ x_k]_i &=&T( T^{-1}( [ x_k]_j;\theta_j);\theta_i).
 \nonumber
\end{eqnarray}
As an example, when $x_k$ is a location on the Cartesian plane, and, $\theta_j$ is the position of sensor $j$, $T$ is given by
\begin{eqnarray}
 T(x_k;\theta_j ) &\triangleq &x_k - \theta_j,
\notag \\ 
 T( T^{-1}( [x_k]_j;\theta_j);\theta_i) &=& [x_k]_j + \theta_j - \theta_i. \notag
\end{eqnarray}
For simplicity in notation, we will denote $T( T^{-1}( .;\theta_j);\theta_i)$ by ${\cal T}_\theta(.)$ when semantics is clear from the context\footnote{Note that, when the calibration parameters also include orientation angles, ${\cal T}_\theta(.)$ involves rotation matrices accordingly.}.

In this section, it is revealed how local filtering distributions are used in the quad-term update. When $\theta$ are respective quantities, these distributions become independent of~$\theta$ because both the state and the measurement variables are in the same local coordinate frame. In other words, at sensor~$j$
\begin{equation}
 p( \vect{X}_k, \tau^j_k | \vect{Z}^j_{1:k}, \theta_j) \equiv p( [ \vect{X}_k]_j, \tau^j_k | \vect{Z}^j_{1:k}),
 \label{eqn:post_reparam}
\end{equation}
holds for the filtering posterior, for any configuration of $\tau^j_k$.

In the prediction stage of filtering, the multi-object transition kernel in~\eqref{eqn:simpmopred} also becomes independent of $\Theta$, so, the Chapman-Kolmogorov equation for finding the prediction density at sensor~$j$ (together with the empirical Bayes selection of association priors during iterations as explained in Section~\ref{sec:problemdef})
becomes
\begin{multline}
p([{\vect X}_k]_j|{\vect Z}_{1:k-1}^j) = \int \pi([{\vect X}_k]_j|[{\vect X}_{k-1}]_j ) \\
\times p([{\vect X}_{k-1}]_j, \tau^j_{k-1}={\bm {\thickbar{\tau}}}^j_{k-1}|{\vect Z}_{1:k-1}^j)\mathrm{d}[{\vect X}_k]_j.
\label{eqn:pred_resparam}
\end{multline}

Note also that the entropy terms in~\eqref{eqn:corolH} that are conditioned on a single sensor's measurements measure the uncertainty of the above densities. Consequently, they also become independent of $\Theta$, i.e., $ H({ \cal X}_k|{  \cal Z}^j_{1:k-1},\Theta )$ equals to $ H({ \cal X}_k|{  \cal Z}^j_{1:k-1} )$ for example, which highlights its relevance to the local prediction accuracy.
\vspace{-6pt}
\subsection{The quad-term update for calibration}
\label{sec:quadtermcalib}
Now, let us expand the quad-term time updates in~\eqref{eqn:quadupdate2}, and, explicitly show the aforementioned simplifications. We start with $s_j^k$ which is the scale factor of the local Bayesian filter at sensor~$j$: 

\begin{eqnarray}
 \mspace{-20mu}s^k_{j}({\vect Z}_k^j,\theta) &= &p_{\tau^j_k}({\vect Z}_k^j|{\vect Z}_{1:k-1}^j,\theta),  \nonumber \\ 
 &=& \!\int\!l_j({\vect Z}_k^j|{\vect X}_k,\tau^j_k,\theta) p({\vect X}_k|{\vect Z}_{1:k-1}^j,\theta)\mathrm{d}{\vect X}_k, \nonumber \\
 &=&\!\int\!l_j({\vect Z}_k^j|[{\vect X}_k]_j,\tau^j_k) p([{\vect X}_k]_j|{\vect Z}_{1:k-1}^j)\mathrm{d}[{\vect X}_k]_j,
 \label{eqn:s_jkUpdate}
\end{eqnarray}
where in the last line independence from $\theta$ is asserted. Because $s_j^k$ does not depend on $\theta$, we denote~$s^k_{j}({\vect Z}_k^j,\theta)$ by $s^k_{j}({\vect Z}_k^j)$ in the rest of the article.

Next, let us consider $r_{ij}^k$ which has terms in different coordinate frames: 
\begin{eqnarray}
 r_{ij}^k({\vect Z}_k^i,\theta) &=&p_{\tau^{i,j}_k}({\vect Z}_k^i| {\vect Z}_{1:k}^j, \theta ) \nonumber \\
&\mspace{-120mu}=&\mspace{-60mu} \!\int\! l_i({\vect Z}_k^i| {\vect X}_k,\tau^{i}_k,\theta)p({\vect X}_k,\tau^j_k|{\vect Z}_{1:k}^j,\theta) \mathrm{d}{\vect X}_k \nonumber \\
&\mspace{-120mu}= &\mspace{-60mu}\!\int\! l_i\left({\vect Z}_k^i| T( {\vect X}_k;\theta_i),\tau^{i}_k\right)
\nonumber p\left(\!T^{-1}({\vect X}_k;\theta_j),\tau^j_k|{\vect Z}_{1:k}^j\right) \mathrm{d}{\vect X}_k \\
&\mspace{-120mu}=&\mspace{-60mu}
\!\int\! l_i\left({\vect Z}_k^i| {\cal T}_\theta( [{\vect X}_k]_j),\tau^{i}_k\right)p\left(\![{\vect X}_k]_j,\tau^j_k|{\vect Z}_{1:k}^j\right) \mathrm{d}[{\vect X}_k]_j.
 \label{eqn:r_ijkUpdate}
\end{eqnarray}
In the third line above, the coordinate transformations are substituted explicitly. The last line follows from that the filtering distribution~\eqref{eqn:post_reparam} is in the $j$th local frame.

\subsection{Evaluation of the quad-term update based on single sensor filtering distributions}
\label{sec:quadtermeval}

{Here, we discuss the evaluation of the quad-term likelihood given local filtering distributions and single sensor association configurations\hfil which \hfil we \hfil denote\hfil for \hfil sensor~$j$  \hfil by}   \\ $p([\vect{X}_k]_j,\tau^j_k=\hat \tau^j_k | \vect{Z}^j_{1:k})$ and $\hat \tau^j_k$, respectively, as explained in Section~\ref{sec:statement}.

Instead of considering evaluation for the most probable association hypothesis ${\bm {\thickbar{\tau}}}^{i,j}_k $ which is infeasible to find, we propose to use the local results $\hat \tau^{i,j}_k \triangleq (\hat \tau^{i}_k, \hat \tau^j_k)$ as a reasonable approximation to this configuration and substitute them in~\eqref{eqn:s_jkUpdate}--\eqref{eqn:r_ijkUpdate}. These approximations can be found regardless of $\theta$ as discussed earlier in this section by using one of the well studied algorithms in the literature~\cite{Vo2015} such as solving a $2-D$ association problem at each time step~\cite{Poore2006} to find $\hat \tau^j_k$. We detail this approach for a linear Gaussian state space model in~Section~\ref{sec:MonteCarloLBP}.

Given these local results and their exchange over the network, one can consider an in-network computation scheme for evaluating~\eqref{eqn:s_jkUpdate}--\eqref{eqn:r_ijkUpdate}. Specifically, these terms (and, the other factors of the quad-term update which are obtained by replacing $i$~and~$j$ in these expressions) can be found at the sensor platform where the measurements to be substituted for evaluation are stored, i.e., sensors $j$ and $i$, respectively, for $s_j^k$ and $r_{ij}^k$. Substitution of the measurement histories on the conditioning side will have been carried out by local filtering. 

More explicitly, $s_j^k$ in~\eqref{eqn:s_jkUpdate} (or, $s_i^k$) becomes a product of similar terms when $\hat \tau^k_j$ is substituted in~\eqref{eqn:s_jkUpdate}, and, its computation is carried out during the local filtering of sensor~$j$'s (or, sensor~$i$'s) measurements using
\begin{eqnarray}
 s^k_{j}({\vect Z}_k^j) &=&\prod_{ o =1 }^M  s^k_{j,o}( z^j_{k,o} )  \label{eqn:s_jo} \\
 s^k_{j,o}( z^j_{k,o})  &\triangleq&  \int\! g_j( z^j_{k,o}| x_k') p_{m'}(x_k'|{\vect Z}_{1:k-1}^j)\mathrm{d} x_k' \nonumber
 \end{eqnarray}
 where $m'={\hat \tau^j_k(o)}$ and the density in the integral is the $m'$th marginal of the local prediction density, the $m$th of which is given by
 \begin{multline}
 \mspace{-20mu}p_m(x'|{\vect Z}_{1:k-1}^j)\triangleq 
 \int p(\vect{X}_k\!=\left[\!{x_{k,1},\ldots,x_{k,m-1},x',x_{k,m+1},}\right.\\
 \left.{x_{k,m+2},\ldots,x_{k,M}}\right],\!\tau^j_k=\!\hat\tau^j_k|\vect{Z}_{1:k-1}^j) \\ \mathrm{d}x_{k,1}\ldots\mathrm{d}x_{k,m-1}\mathrm{d}x_{k,m+1}\ldots\mathrm{d}x_{k,M}.
 \nonumber
\end{multline}

The term $r_{ij}^k$ in~\eqref{eqn:r_ijkUpdate} (or, $r_{ji}^k$) is also computed based on these local filtering distributions. The integration in the RHS of~\eqref{eqn:r_ijkUpdate} implicitly assumes that the ordering of individual objects in the local multi-object vectors are the same. In a networked setting, however, this is not necessarily the case and the identities of the fields in the state vector may differ~\cite{Guibas2008}. In order to tackle with this unknown correspondance, we introduce an additional permutation random variable $\gamma_k$ (see, e.g.,~\cite{Huang2009}) for relating the fields of a multi-object vector ${\vect X}_k$ as ordered locally at sensor~$i$ and $j$, such that the $m$th field of the state vector at sensor~$i$ refers to the same object in the $\gamma_k(m)$th field of the state vector at sensor~$j$. For example, 
\begin{equation}
 [ x_{k,m} ]_i = {\cal T}_\theta( [ x_{k,\gamma_k(m) } ]_j ). \notag
\end{equation}

Suppose that an estimate $\hat \gamma_k$ of this quantity is provided. After substituting in~\eqref{eqn:r_ijkUpdate} together with $\hat \tau^{i,j}_k $ one obtains
\begin{eqnarray}
 r_{ij}^k({\vect Z}_k^i , \theta) &=& \prod_{o=1}^M r^k_{ij,o}( z_{k,o}^i ,\theta), \label{eqn:rij_o} \\
 r^k_{ij,o}( z_{k,o}^i ,\theta) & \triangleq &\int\! g_i\left( z_{k,o}^i| {\cal T}_\theta( x_k' )\right)
 p_{m' }(\!x_k'|{\vect Z}_{1:k}^j) \mathrm{d}x_k' \nonumber 
 \end{eqnarray}
where $m'= \hat \gamma_k( \hat \tau^i_k(o)) $ and the density inside the integral is the $m'$th marginal of the filtering distribution local to sensor~$j$. In Appendix~\ref{sec:mlgamma}, we show that \eqref{eqn:rij_o} replaces the likelihood for $\gamma_k$ when $m'= \gamma_k( \hat \tau^i_k(o))$ and an ML estimate $\hat \gamma_k$ can be found in a way similar to solving the data association problem in local filtering, which is detailed later in Section~\ref{sec:ssfilter}.
   
Finally, the scale factor~\eqref{eqn:quadscale} is computed. This involves finding the measurement distributions in~\eqref{eqn:quadscale} using the prediction distribution in~\eqref{eqn:pred_resparam} for both sensors~$i$~and~$j$ together with their likelihoods. This leads to the following two decomposition: The first term in~\eqref{eqn:quadscale} is found as
\begin{eqnarray}
p( {\vect{Z}^i_k}', {\vect{Z}^j_k}' | \vect{Z}^i_{1:k-1}, \theta   ) &=& \prod_{o=1}^M p_o( 
{{z^i}'}_{\!\!\!\!k,o}, {{z^j}'}_{\!\!\!\!k,\xi(o)}|\vect{Z}^i_{1:k-1}, \theta) \nonumber \\
p_o( {z^i}, {z^j}|\vect{Z}^i_{k-1}, \theta) &\triangleq& \int g_i( {z^i} |x_k') p_{m'}(x_k'|\vect{Z}^i_{1:k-1}) \mathrm{d}x_k'  \nonumber \\
&&\mspace{-100mu} \times \int g_j( {z^j}| {{\cal T}_\theta}^{-1}(x_k') )p_{m'}(x_k'|\vect{Z}^i_{1:k}) \mathrm{d}x_k' 
\label{eqn:p_oGhi}
\end{eqnarray}
where $\xi(o) = { {{\hat\tau}^{j -1}_k}} \circ \hat \tau^i_k(o) $ maps the $o$th measurement at sensor~$i$ to the corresponding one in sensor~$j$, and, $m'=\hat \tau^i_k(o)$ in the second line.

The second term in~\eqref{eqn:quadscale} is found as
\begin{eqnarray}
p( {\vect{Z}^i_k}',{\vect{Z}^j_k}' | \vect{Z}^j_{1:k-1}, \theta   ) &=&\prod_{o=1}^M  p_o({{z^i}'}_{\!\!\!\!k,o}, {{z^j}'}_{\!\!\!\!k,\xi(o)}|\vect{Z}^j_{k-1}, \theta) \nonumber \\
p_o(z^i, z^j|\vect{Z}^j_{k-1}, \theta) &\triangleq& \int g_j(z^j| x_k' ) p_{m'}(x_k'|\vect{Z}^j_{1:k-1})\mathrm{d}x_k'\nonumber \\
&& \mspace{-100mu} \times \int g_i(z^i| {{\cal T}_\theta}(x_k') )   p_{m'}(x_k'|\vect{Z}^j_{1:k})\,\, \mathrm{d}x_k'
\label{eqn:p_oGhj}
\end{eqnarray}
where $m'= \hat \gamma_k \circ \hat \tau^i_k(o)$ in the last line is the object that corresponds to the $o$th measurement at sensor~$i$.

Consequently, the scale factor is found as 
\begin{eqnarray}
 \kappa_k(\theta) &= & \prod_{o=1}^M \kappa_{k,o}(\theta) 
 \label{eqn:qsdecomp} \\
 \kappa_{k,o}(\theta) &\triangleq& \int \left( p_o( {z^i}, {z^j}|\vect{Z}^i_{1:k-1}, \theta) \right. \nonumber \\
 && \left. \mspace{50mu} \times p_o({z^i},{z^j}|\vect{Z}^j_{1:k-1}, \theta) \right)^{1/2} \mathrm{d}{z^i} \mathrm{d}{z^j}, \nonumber
\end{eqnarray}
using the densities found in~\eqref{eqn:p_oGhi}~and~\eqref{eqn:p_oGhj}.

The expressions above describe the evaluation of the quad-term likelihood in terms of single sensor filtering distributions that can be obtained using any filtering algorithm with individual measurement histories. The use of the local filtering distributions provides scalability with the number of sensors for parameter estimation in the state space model in~Section~\ref{sec:problemdef}. 

These computations can be distributed in the pair $(i,j)$ as follows: Both sensors~$i$~and~$j$ perform local filtering and exchange the resulting posterior densities at every step, as well as $s_i$ and $s_j$, respectively, found using~\eqref{eqn:s_jo}. Based on the received densities, sensors $i$ and~$j$ evaluate $r_{ij}$ and $r_{ji}$, respectively, using~\eqref{eqn:rij_o}. As part of the filtering process, they realise the Chapman-Kolmogorov equation in~\eqref{eqn:pred_resparam} with their local posterior, as well as the remote posterior recently received. These densities are then used in~\eqref{eqn:p_oGhi},~\eqref{eqn:p_oGhj},~and,~\eqref{eqn:qsdecomp} to compute the scale factor for the next step. The scale factor hence found in the previous step, therefore, is substituted in~\eqref{eqn:quadupdate2}~together with the four terms computed. This is repeated for~$k=1,\ldots,t$, and, the quad-term likelihood~\eqref{eqn:quadpotwnorm} is computed, as a result.


\section{A Monte Carlo LBP algorithm for sensor calibration in linear Gaussian state space models}
\label{sec:MonteCarloLBP}

In this section, we consider a linear Gaussian state space (LGSS) model within the probabilistic graphical model in~\figurename~\ref{fig:distmodel}, and, specify an algorithm for estimation of $\theta$s that combines the quad-term calibration likelihood evaluation detailed in Section~\ref{sec:quad4calibration} with BP message passing on the resulting pairwise MRF model (Section~\ref{sec:MRF}). This algorithm uses Monte Carlo methods for realising BP~\cite{Sudderth2010} and facilitates scalability by building upon single sensor filtering as required by the quad-term approximation. As a result, an efficient inference scheme over the model in~\figurename~\ref{fig:distmodel} is achieved.

The state space model we consider is specified by a linear state transition with process noise that is additive and Gaussian, and, linear measurements with independent Gaussian measurement noise,~i.e.,
\begin{eqnarray}
 \pi( x_{k}|x_{k-1}) &= &{\cal N}( x_{k} ; \vect{F}x_{k-1}, \vect{Q})  \label{eqn:lgtrans} \\
 g_j( z_k^j | x_k; \theta_j ) &= &{\cal N}( z_k^j; \vect{H}_j [x_k]_j , \vect{R}_j ) \label{eqn:lgobs}
\end{eqnarray}
for $j=1,\ldots,N$, where ${\cal N}(.;{\mu}, {\vect P})$ is a multi-dimensional Gaussian density with mean vector $\mu$ and covariance matrix~${\vect{ P}}$.

Here, $x_k$ is the concatenation of position and velocity (on a $2-D$ Euclidean plane, without loss of generality). The matrices $\vect{F}$ and  $\vect{Q}$ model motion with unknown acceleration (equivalently, manouevres), and, are selected as
\begin{eqnarray}
 \vect{F} =   \left[ \begin{array}{rr} \vect{I},& \Delta T \times \vect{I} \\
   \vect{0},&\vect{I} \end{array} \right] , \, \, \vect{Q} = \sigma^2 \left[ \begin{array}{rr} q_{1}\vect{I},&q_{2}\vect{I} \\
   q_{2}\vect{I},& q_{3}\vect{I} \end{array} \right]
   \notag
\end{eqnarray}
where $\vect{I}$ and $\vect{0}$ are the $2\times2$ identity and zero matrices, respectively. $\Delta T$ is the time difference between consecutive steps. $\vect{Q}$ is positive definite and parameterised with $\sigma^2$, and, $0<q_1<q_2<q_3<1$ specifying the magnitude of the uncertainty, and, contributions of higher order terms, respectively~\footnote{It can easily be shown that this state transition model is invariant under the selected coordinate frame for~$x_k$s.}.

In the measurement model, $\vect{R}_j$ is the measurement noise covariance, and, $\vect{H}_j$ is the observation matrix which we assume forms an observable pair with $\vect{
F}$ (e.g., $\vect{H}_j = \left[ \vect{I}, \vect{0} \right]$).
\subsection{Local single sensor filtering}
\label{sec:ssfilter}
We now focus on filtering and provide explicit formulae that adopts the recursions in \eqref{eqn:update}--\eqref{eqn:factcent_emp} for a single sensor in the LGSS model. We use the empirical Bayes approach explained in Section~\ref{sec:statement} for scaling with time under data association uncertainties. This approach  corresponds to the single frame data association solution in multi-object tracking~\cite{Poore2006}. 

{First, let us consider the prediction stage at time $k$ in which we are given a filtering density evaluated at the most likely data association hypothesis $\hat \tau^j_{k-1}$ of the previous step~\footnote{Note that the empirical prior on $\tau^j_{k-1}$ is selected as in~\eqref{eqn:empiricalprior} leading to this posterior be identically zero for all values of $\tau^j_{k-1}$ other than $\hat \tau^j_{k-1}$.}. The latter is a product of its marginals}

\begin{multline}
 p( \vect{X}_{k-1}, \tau^j_{k-1}=\hat\tau^j_{k-1} |\vect{Z}^j_{1:k-1}) = \nonumber \\
 \mspace{40mu} \prod_{m=1}^M p_m( x_{k-1,m}, \tau^j_{k-1}=\hat\tau^j_{k-1} |\vect{Z}^j_{1:k-1})
\end{multline}
where
\begin{equation}
p_m( x_{k-1,m}, \tau^j_{k-1}=\hat\tau^j_{k-1} |\vect{Z}^j_{1:k-1})= p( x_{k-1,m}|z^{j,m}_{1:k-1}).\notag
\end{equation}
Here, $z^{i,m}_{1:k-1}$ denotes the measurements induced by object $m$ from step $1$ to $k-1$, i.e.,
\begin{equation}
 z^{j,m}_{1:k-1} \triangleq \left( z^{j }_{k-1,{\rho^j_{k-1}(m)}},z^{j}_{k-2,{\rho^j_{k-2}(m)}},\ldots,z^{j}_{1,{\rho^j_{1}(m)}} \right),
 \nonumber
\end{equation}
where $\rho$ is the inverse of $\tau$, i.e., $\rho \circ \tau $ is the identity permutation.

Consequently, the $m$th marginal of the posterior at $k-1$ is a Gaussian density that can be obtained equivalently by Kalman filtering~\cite{Ristic2004} over $z^{j,m}_{1:k-1}$, i.e., 
\begin{equation}
p_m( x_{k-1,m}|z^{j,m}_{1:k-1})
 ={\cal N}(x_{k-1,m};  {\hat x^j_{k-1,m}}, \vect{P}^j_{k-1,m}),
 \label{eqn:postmarginal}
\end{equation}
with the mean and covariance matrices over time specifying the $m$th ``track.''

Hence, the prediction density in~\eqref{eqn:pred_resparam} (Section~\ref{sec:quadtermcalib}) evaluated at $\tau^j_{k-1}=\hat \tau^j_{k-1}$ for the state transition in~\eqref{eqn:lgtrans} is given~by
\begin{eqnarray}
 p(\vect{X}_{k}|\vect{Z}^j_{1:k-1} ) &=& \prod_{m=1}^M {\cal N}(x_{k,m};{\hat x}^j_{k|k-1,m},\vect{P}^j_{k|k-1,m} ) \label{eqn:lgpred} \\
 {\hat x}^j_{k|k-1,m} &= &\vect{F} {\hat x^j_{k-1,m}}, \nonumber \\
 \vect{P}^j_{k|k-1,m} &=& \vect{F}\vect{P}^j_{k-1,m}\vect{F}^T + \vect{Q}, \nonumber
\end{eqnarray}
where the last two lines are Kalman prediction equations with $\vect{(.)}^T$ denoting matrix transpose.

Next, let us consider the update stage in which we use the prediction density~\eqref{eqn:lgpred} with $M$ measurements concatenated in $\vect{Z}_k^j$ and the measurement likelihood. This likelihood is found by substituting~\eqref{eqn:lgobs}~in~\eqref{eqn:simpmolhood}. First, we use this term within the likelihood for the association variable $\tau^j_k$ which --as it is mutually independent from $\tau^j_1,\ldots,\tau^j_{k-1}$-- is given~by
\begin{eqnarray}
 l_j(\vect{Z}^j_{1:k}|\tau^j_k)&\propto& p_{\tau^j_k}(\vect{Z}^j_{k}|\vect{Z}^j_{1:k-1}) \label{eqn:lgassoclhood} \\
 &=& \int l(\vect{Z}_k^j|\vect{X}_{k},\tau^j_k )p(\vect{X}_{k}|\vect{Z}^j_{1:k-1})\mathrm{d} \vect{X}_k \nonumber \\
 &=&  \prod_{o=1}^M \int g_j(z_{k,o}^j|x'_k) p_{\tau^j_k(o)}( x_k'|\vect{Z}^j_{1:k-1}) \mathrm{d}x_k' \nonumber \\
 &=&  \prod_{o=1}^M \int g_j(z_{k,o}^j|x_k') p( x_k'|z^{j,\tau^j_k(o)}_{1:k-1}) \mathrm{d}x_k'. \nonumber
\end{eqnarray}
The first line above follows from that the joint distribution of $Z^j_{1:k-1}$ and $\tau^j_k$ is independent of the latter.

The prior distribution for $\tau^j_k$ is non-informative as given in~\eqref{eqn:equalikelyprior}, so, the ML estimate using \eqref{eqn:lgassoclhood} coincides with the MAP estimate and it is given~by 
\begin{equation}
 \hat \tau^j_k = \arg \max_{{\tau^j_k}\in {\cal S}_M } l_j(\vect{Z}^j_{1:k}|\tau^j_k)
 \label{eqn:mlproblem}
\end{equation}
where ${\cal S}_M$ is the set of $M$-permutations.

An equivalent problem is found by taking the logarithm of the objective function in the combinatorial optimisation problem above as follows:
\begin{eqnarray}
  \hat \tau^j_k&=& \arg \max_{  {\tau^j_k}\in {\cal S}_M } \sum_{o=1}^M  c\big(o,m=\tau^j_k(o)\big) \label{eqn:assoc} \\
 c(o,m) &\triangleq & \log \int g_j(z_{k,o}^j|x_k') p( x_k'|z^{j,m}_{1:k-1}) \mathrm{d}x_k', \nonumber
\end{eqnarray}
for $o,m=1,\ldots,M$.

This cost for the LGSS model is explicitly found using the prediction distribution~\eqref{eqn:lgpred} and the measurements within the KF innovations~\cite{Ristic2004}~as
\begin{eqnarray}  
  c(o,m)  &=& \log {\cal N}( z^j_{k,o}; {\hat z}^j_{k,m}, \vect{S}^{j}_{k,m} ) \label{eqn:cost}\\
 \hat z^j_{k,m} &= &{\vect{H}_j}{\hat x}^j_{k|k-1,m}, \,\,\,
  {\vect{S}^j_{k,m} = \vect{R}_j+\vect{H}_j\vect{P}^j_{k|k-1,m}\vect{H}^T_j}. \nonumber
\end{eqnarray}

The optimisation in~\eqref{eqn:assoc} is a $2-D$ assignment problem which can be solved in polynomial time with $M$ (despite that the search space ${\cal S}_M$ has a factorial size) using one of the well known solvers~\cite{Burkard2012} including the auction algorithm~\cite{Bertsekas1988}. This algorithm operates over a matrix of costs obtained by $\vect{C}=[c(o,m)]$ to iteratively find the $M$ pairs corresponding to the best permutation $\hat \tau^j_{k}$ in the ML problem~\eqref{eqn:mlproblem}. Here, computation of the $M^2$ cost matrix usually has the predominant computational time.

Next, we consider the state distribution update (see \eqref{eqn:conditionedposterior} in Section~\ref{sec:statement}) and assert the empirical (model) prior in \eqref{eqn:empiricalprior}. As a result, the filtering density at $k$ becomes a product of its marginals each of which is a Gaussian as in~\eqref{eqn:postmarginal} found by the KF update~\cite{Ristic2004}, i.e., 
\begin{equation}
  p(\vect{X}_{k}, \tau^j_k = \hat \tau^j_k |\vect{Z}^j_{1:k} )=\prod_{m=1}^M{\cal N}(x_{k,m}; \hat x_{k,m}, \vect{P}_{k,m}) 
  \label{eqn:localpost_lgss}
\end{equation}
where,
\begin{eqnarray}
 \hat x_{k,m} &=& \hat x_{k|k-1,m} +\vect{K}_{k,m}(z^j_{k,\rho_k^j(m)} - \vect{H}_j\hat x_{k|k-1,m}  ) \nonumber \\
 \vect{P}_{k,m}&=& \left(\vect{I}-\vect{K}_{k,m}\vect{H}_j \right)\vect{P}^j_{k|k-1,m} \nonumber \\
\vect{K}_{k,m} &\triangleq&  \vect{P}^j_{k|k-1,m}\vect{H}_j^{T}{{\vect{S}^j_k}}^{-1}_{,m}.\nonumber
\end{eqnarray}
Note that, because the posterior density is now non-zero only for $\tau^j_k = \hat \tau^j_k$, the marginalisation over $\tau^j_k$ (see, e.g.,~\eqref{eqn:pred}) in the following prediction stage reduces to~\eqref{eqn:pred_resparam} and~\eqref{eqn:lgpred}.

\subsection{Evaluation of the calibration quad-term in the LGSS Model}
\label{sec:quadtermevalLGSS}
Let us consider the evaluation of $s_j^k$ given by~\eqref{eqn:s_jo} using the formulae for the LGSS model introduced in Section~\ref{sec:ssfilter}. By comparison with the cost term in~\eqref{eqn:assoc}~and~\eqref{eqn:cost}, it can easily be seen that $s_j^k$ is the exponential of the association cost for $\hat \tau^j_k$, i.e.,
\begin{equation}
 s_j^k( \vect{Z}^j_k ) = \exp \sum_{o=1}^M c\big(o, \hat \tau^j_k(o) \big).
 \label{eqn:s_jkeval}
\end{equation}

Next, let us consider~\eqref{eqn:rij_o} for evaluating $r_{ij}^k$. Evaluation of this term involves finding the object identity correspondance $\hat \gamma_k$ by solving a $2$-D assignment as explained in Appendix~\ref{sec:mlgamma}. In the LGSS model, the assignment cost matrix ${\cal D}=[d(o,m)]$ is found as
\begin{eqnarray}
 d(o,m) &= &\log {\cal N}(z^i_{k,o}; \hat z^{i}_{k,m}, {\vect S}^i_{k,m} ) \label{eqn:d_lgss}\\
  \hat z^i_{k,m} &= &{\vect{H}_i}{\cal T}_\theta( {\hat x}^j_{k,m} ), \nonumber \\
  {\vect{S}^i_{k,m}} &= &{\vect{R}_i+\vect{H}_i{\cal T}_\theta( \vect{P}^j_{k,m})\vect{H}^T_i}. \nonumber
\end{eqnarray}
for $o,m=1,\ldots,M$. Here, the second order statistics $ \vect{P}^j_{k,m}$ is also transformed by applying any rotations involved in ${\cal T}_\theta$ to its eigenvectors. The best assignment which here encodes $\hat \gamma_k$ is found using the auction algorithm~\cite{Bertsekas1988} as well, similar to the assignment in the Bayesian filtering update (Sec.~\ref{sec:ssfilter}). Using this estimate, the quad-term factor is computed using
\begin{equation}
 r_{ij}^k( \vect{Z}^i_k, \theta ) = \exp \sum_{o=1}^M d\Big(o, \hat \gamma_k\big( \hat \tau^i_k(o) \big) \, \Big).
 \label{eqn:r_ijkeval}
\end{equation}

In order to evaluate the other factors of the quad-term update, i.e., $s^i_k$ and $r_{ji}^k$, similar computations are used. It suffices to replace $i$ in the subscripts/superscripts of the expressions above with $j$, and, vice versa.

Finally, let us consider the scale factor in~\eqref{eqn:qsdecomp}. Let us use the notation introduced in the previous section for expressing the densities inside the integration. Starting with~\eqref{eqn:p_oGhi}, one obtains
\begin{eqnarray}
  p_o( {z^i}, {z^j}|\vect{Z}^i_{1:k-1}, \theta) &= &p( {z^i}, {z^j}|{z}^{i,\hat \tau^i_{k-1}(o) }_{1:k-1}, \theta) \label{eqn:p_ozlgss} \\
  &=& {\cal N}( [z^i,z^j]^T; {\mu}_1, \vect{\Sigma}_1 )   \nonumber \\
  {\mu}_1&=& \left[ {\begin{array}{c}
  \hat z^i_{k,m} \\
   {\vect H}_j {\cal T}^{-1}_\theta(\hat x^i_{k, m } )  \\
  \end{array} }  \right] \nonumber \\
  \vect{\Sigma}_1 &=& \left[  {\begin{array}{lr}
   \vect{S}^i_{k,m} & \vect{0}   \\
   \vect{0} & \vect{R}_j + \vect{H}_j {\cal T}^{-1}_\theta(\vect{P}^i_{k,m} )\vect{H}_j^T \end{array} }     \right] \nonumber 
\end{eqnarray}
where $m=\hat \tau^i_{k-1}(o)$, and, $\hat z^i_{k,m}$ and $\vect{S}^i_{k,m}$ are computed as in~\eqref{eqn:cost} with $i$ substituted in place of~$j$.

The second density in~\eqref{eqn:qsdecomp} conditioned on sensor~$j$'s history, i.e., \eqref{eqn:p_oGhj},  is similarly found as
\begin{eqnarray}
 p_o( {z^i}, {z^j}|\vect{Z}^j_{1:k-1}, \theta) &=& {\cal N}( [z^i,z^j]^T; {\mu}_2, \vect{\Sigma}_2 )   \nonumber \\
  {\mu}_2&=& \left[ {\begin{array}{c}
  {\vect H}_i {\cal T}_\theta(\hat x^j_{k, m } ) \\
     \hat z^j_{k,m}  \\
  \end{array} }  \right] \nonumber \\
  \vect{\Sigma}_2 &=& \left[  {\begin{array}{lr}
    \vect{R}_i + \vect{H}_i {\cal T}_\theta(\vect{P}^j_{k,m}) \vect{H}_i^T & \vect{0}   \\
   \vect{0} & \vect{S}^j_{k,m}  \end{array} }     \right] \nonumber 
\end{eqnarray}
where $m=\hat \gamma_{k-1} \circ \hat \tau^i_{k-1}(o)$ and, $\hat z^j_{k,m}$ and $\vect{S}^j_{k,m}$ are given in~\eqref{eqn:cost}.

Using the densities above and integration rules for Gaussians, the $o$th term of the scale factor in~\eqref{eqn:qsdecomp} is found as
\begin{multline}
\mspace{-20mu} \kappa_{k,o}(\theta) = \label{eqn:K_koeval}\\
\frac{ \left( \left| \vect{\Sigma}^{-1}_1 \right|\left| \vect{\Sigma}^{-1}_2 \right|\right)^{1/4} }{ \left| \frac{ \vect{\Sigma}^{-1}_1 + \vect{\Sigma}^{-1}_2}{2} \right|^{1/2}  }\exp \left\{ -\frac{1}{4}\left( \mu_1^T \vect{\Sigma}_1^{-1}\mu_1 + \mu_2^T\vect{\Sigma}_2^{-1}\mu_2 \right) \right.\\
\left.+\frac{1}{4}\left( \vect{\Sigma}_1^{-1}\mu_1 + \vect{\Sigma}_2^{-1}\mu_2 \right)^T \left( \vect{\Sigma}_1^{-1}+\vect{\Sigma}_2^{-1} \right)^{-1}\right. \\
\left. \times \left(\vect{\Sigma}_1^{-1}\mu_1 + \vect{\Sigma}_2^{-1}\mu_2\right) \right\}.
\end{multline}

In a distributed setting, the scale factor expressions above are computed both at sensors $i$~and~$j$, for which the prediction stage given in~\eqref{eqn:lgpred} is carried out for both the local posterior and the posterior recevied from the other sensor at time~$k-1$.

\subsection{Sampling from the calibration marginals using non-parametric BP}


In this section, we introduce particle based representations and Monte Carlo computations~\cite{Casella2005} for the realisation of (loopy) BP message passings. Note that Sections~\ref{sec:ssfilter}~and~\ref{sec:quadtermevalLGSS} specify the evaluation of edge potentials given in \eqref{eqn:quadpotwnorm} for $(\tau^{i}_{1:t},\tau^{j}_{1:t})=(\hat\tau^{i}_{1:t},\hat\tau^{j}_{1:t})$.

For sampling from the marginal parameter posteriors, we adopt the approach detailed in~\cite[Sec.VI]{Uney2016} for carrying out LBP belief update and messaging in~\eqref{eqn:bpstate}~and~\eqref{eqn:bpmess}, respectively. Given $L$ equally weighted samples from $\tilde p_i(\theta_i)$,~i.e.,
\begin{equation}
 \theta_i^{(l)} \sim  \tilde p_i(\theta_i),
 \label{eqn:locSamples}
\end{equation}
for $l=1,\dots,L$, the edge potentials are evaluated to obtain 
\begin{equation}
 \psi_{ij}(\theta_i^{(l)},\theta_j^{(l)})=\tilde l\left({\vect{Z}^i_{1:t},\vect{Z}^j_{1:t}|\theta=(\theta_i^{(l)},\theta_j^{(l)})}\right).
 \label{eqn:epot_eval}
\end{equation}

Consider the BP message from node $j$ to $i$ in~\eqref{eqn:bpmess}. Suppose that independent identically distributed (i.i.d.) samples from the (scaled) product of the $j$th local belief and the incoming messages from all neighbours except $i$ are given, i.e.,
\begin{equation}
 \bar \theta_{j}^{(l)} \sim \tilde p_{j}(\theta_j) \prod_{i' \in ne(j)/i} m_{i'j} (\theta_j)\,\,\,\text{for}\,\,\,l=1,...,L.
 \label{eqn:prodsamples1}
\end{equation}

These samples are used with kernel approximations in order to represent the message from node $j$ to~$i$ (scaled to one), in the NBP approach~\cite{Sudderth2010}. We use Gaussian kernels leading to the approximation given by
\begin{eqnarray}
 \hat m_{ji}(\theta_i) &=&\sum_{l=1}^L \omega_{ji}^{(l)} {\cal N}( \theta_i ; \theta_{ji}^{(l)}, \Lambda_{ji}),
  \label{eqn:bpmessemp} \\
   \theta_{ji}^{(l)} &=& T( T^{-1} (\bar \theta_{j}^{(l)};\theta_j^{(l)}); \theta_{i}^{(l)}), \notag \\
 \omega_{ji}^{(l)} &=&\frac{ \psi_{i,j}( \theta_i^{(l)}, \theta_j^{(l)} )}{\sum_{l'=1}^L  \psi_{i,j}(  \theta_i^{(l')}, \theta_j^{(l')} )},\notag 
\end{eqnarray}
where the kernel weights are the normalised edge potentials. $\Lambda_{ji}$ is related to a {\it bandwidth} parameter that can be found using Kernel Density Estimation (KDE) techniques. In particular, we use the rule-of-thumb method in~\cite{Silverm1986} and find
\begin{eqnarray}
 \Lambda_{ji} &=& \left( \frac{4}{(2d+1)L} \right)^{2/(d+4)}  \hat {\vect{C}}_{ji}, \notag \\
 \hat {\vect{C}}_{ji}&=& \sum_{l'}\sum_{l} \omega_{ji}^{(l')}\omega_{ji}^{(l)}(\theta_{ji}^{(l')}-\hat{ \vect{m}}_{ji})(\theta_{ji}^{(l)}-\hat{\vect{m}}_{ji})^T, \notag \\
 \hat  {\vect{m}}_{ji} &= &\sum_{l=1}^L \omega_{ji}^{(l)} \theta_{ji}^{(l)} \notag
\end{eqnarray}
where $\hat {\vect{m}}_{ji}$ and $\hat {\vect{C}}_{ji}$ are the empirical mean and covariance of the samples, respectively, and $d$ is the dimensionality of $\theta_{ji}$s.

Given these messages, let us consider sampling from the updated marginal in \eqref{eqn:bpstate}. We use the weighted bootstrap (also known as sampling/importance resampling)~\cite{Smith1992} with samples generated from the (scaled) product of Gaussian densities with mean and covariance found as the empirical mean and covariance of the particle sets, respectively. In other words, given $\hat  {\vect{m}}_{ji}$ and $\hat {\vect{C}}_{ji}$ as above, we generate
\begin{eqnarray}
\theta_{i}^{(l)} &\sim &f(\theta_i),\,\,\,\ l=1,\ldots,L, \notag \\
f( \theta_i ) &\propto &{\cal N}(\theta_i; \hat {\vect{m}}_{i}, \hat {\vect{C}}_{i})\prod_{j \in ne(i)} {\cal N}(\theta_i; \hat {\vect{m}}_{ji}, \hat {\vect{C}}_{ji}). \notag
\end{eqnarray}

The particle weights for these samples to represent the updated marginal is given by
\begin{eqnarray}
 \omega_{i}^{(l)} &= &\hat \omega_{i}^{(l)}/\sum_{l'=1}^L \hat \omega_{i}^{(l')} \notag \\ 
 \hat \omega_{i}^{(l)} &= & \Big( p_{0,i}(\theta_{i}^{(l)}) \prod_{j \in ne(i)} \hat m_{ji}(\theta_{i}^{(l)})  \Big)/f(\theta_{i}^{(l)}) \notag
\end{eqnarray}
where $p_{0,i}$ is the prior density selected for $\theta_i$ (and, the node potential in~\eqref{eqn:approxpost}). Thus, the local calibration marginal is estimated by
\begin{equation}
 \hat P_{i}(\mathrm{d}\theta_i) = \sum_{l=1}^L \omega_{i}^{(l)} \delta_{\theta_{i}^{(l)}}(\mathrm{d}\theta_i).
 \label{eqn:estmarginal}
\end{equation}

As the final step of the bootstrap, $\{\theta_{i}^{(l)}, \omega_{i}^{(l)} \}_{l=1}^M$ is resampled (with replacement) leading to equally weighted particles from $\tilde p_{i}(\theta_i)$, i.e., $\{ \theta_{i}^{(l)} \}_{l=1}^L$. We follow similar bootstrap steps in order to generate the samples in~\eqref{eqn:prodsamples1}.

After nodes iterate the BP computations described above for $S$ times, each node estimates its location by finding the empirical mean of $\{ \theta_{i}^{(l)} \}_{l=1}^L$. These steps are summarised in Algorithm~\ref{algnbp}.

\begin{algorithm}[t!]
\caption{Pseudo-code for estimation of $\theta$ using the quad-term separable likelihood within Belief Propagation.}
\label{algnbp}
\begin{spacing}{1.5}
{\fontsize{8}{9}\selectfont
\begin{algorithmic}[1]
\algrenewcommand\algorithmicindent{0.75em}
\ForAll{$j \in {\cal V}$} \Comment{{Local filtering}}
\For{$k=1,\dots,t$}
\State {Find} $p(X_k, \tau^j_k = \hat \tau^j_k|\vect{Z}^j_{1:k})$ in~\eqref{eqn:localpost_lgss} as described in Section~\ref{sec:ssfilter}
\State {Find} $s^k_j(\vect{Z}^j_{k})$ {
in~\eqref{eqn:s_jkeval}}
\EndFor
\EndFor
\ForAll{$j \in {\cal V}$} \Comment{{Sample from priors}}
\State Sample $\theta_i^{(l)}\sim p_{0,i}(\theta_i)$ for $l=1,\dots,L$ as in~\eqref{eqn:locSamples}
\EndFor
\For{$s=1,...,S$} \Comment{$S$-steps of LBP}
\ForAll{$(i,j) \in {\cal E}$} \Comment{{Evaluate edge potentials}}
\For{$l=1,\dots,L$}
\State {Find} $ r_{ij}^k( \vect{Z}^i_k, \theta = (\theta_i^{(l)},\theta_j^{(l)}))$ {using \eqref{eqn:d_lgss},~\eqref{eqn:r_ijkeval} for $k=1,\dots t$} 
\State {Find} $r^k_{ji}( \vect{Z}^j_k, \theta = (\theta_i^{(l)},\theta_j^{(l)}))$ for $k=1,\dots t$
\State {Find} $\kappa_{k}( \theta = (\theta_i^{(l)},\theta_j^{(l)}) )$ using \eqref{eqn:qsdecomp},~\eqref{eqn:p_ozlgss}--\eqref{eqn:K_koeval} for $k=1,\ldots,t$
\State {Find} $q( \vect{Z}^i_k, \vect{Z}^j_k |\vect{Z}^i_{1:k-1}, \vect{Z}^j_{1:k-1} ,\theta = (\theta_i^{(l)},\theta_j^{(l)}))$ using \eqref{eqn:quadupdate2} for $k=1,\ldots,t$
\State {Find} $\psi_{i,j}(\theta_i^{(l)},\theta_j^{(l)})$ { in \eqref{eqn:epot_eval}~as the quad-term likelihood in~\eqref{eqn:quadpotwnorm}} 
\EndFor
\EndFor
\ForAll  {$(i,j) \in {\cal E}$} \Comment{{Find LBP message}}
\State Find the kernel representation $\hat m_{ji}(\theta_i)$ in \eqref{eqn:bpmessemp} 
\EndFor
\ForAll  {$i \in {\cal V}$}  \Comment{Update local marginals}
\State Find the updated $\hat P_i$ in~\eqref{eqn:estmarginal} and sample $\theta_i^{(l)} \sim \tilde p_i(\theta_i)$ 
\State $\hat \theta_i \gets  \frac{1}{L} \sum_{l=1}^L \theta_i^{(l)}$
\EndFor
\EndFor
\end{algorithmic}
}
\end{spacing}
\end{algorithm}
\section{Example: Self-localisation in LGSS models}
\label{sec:Example}

In this example, we demonstrate the quad-term node-wise separable likelihood in sensor self-localisation. The LGSS model given by \eqref{eqn:lgtrans}~and~\eqref{eqn:lgobs} is used with process noise parameters selected as $\sigma = 0.5$, $q_1 = 1/4$, $q_2=q_3=1/2$, $q_4=1$. The measurement model for sensor $i$ is given by $\vect{H}_i = [\vect{I}, \vect{0}]$ and $\vect{R}_i = \sigma_n^2 \vect{I}$ with $\sigma_n = 10$ modelling noisy position measurements in the local coordinate frame.

\begin{figure}[t!]
  \centerline{
    \includegraphics[width=\linewidth]{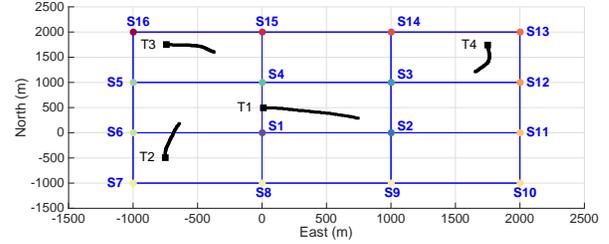} 
  }
  \vspace{-10pt}
  \caption[Example scenario]{Example scenario: 16 Sensors collect measurements from 4 objects (T1-T4) with association uncertainties. Initial positions of the objects are denoted by black squares. Trajectories for $60$ time steps are depicted. The blue lines depict the edges of the MRF model used for estimation.}
  \label{fig:scenario}
  \vspace{-10pt}
\end{figure}

Let us consider the multi-object multi-sensor scenario depicted in \figurename~\ref{fig:scenario}. 16 sensors observe 4 objects moving with data association uncertainties. The locations of the sensors are to be estimated with respect to sensor $1$ which is selected as the origin of the network coordinate system. Therefore $\theta=[\theta_1,\ldots,\theta_{16}]$ with the prior distribution for $\theta_1$ selected as Dirac's delta, i.e., $p_{0,1}(\theta_1)=\delta(\theta_1)$. For the other nodes, the localisation prior, i.e., $p_{0,i}(\theta_i)$ for $i=2,\ldots,16$, is a uniform distribution over the sensing region.

\begin{figure}[b!]
\vspace{-5pt}
  \centering{
  \begin{minipage}{\linewidth}
    \includegraphics[width=0.48\linewidth]{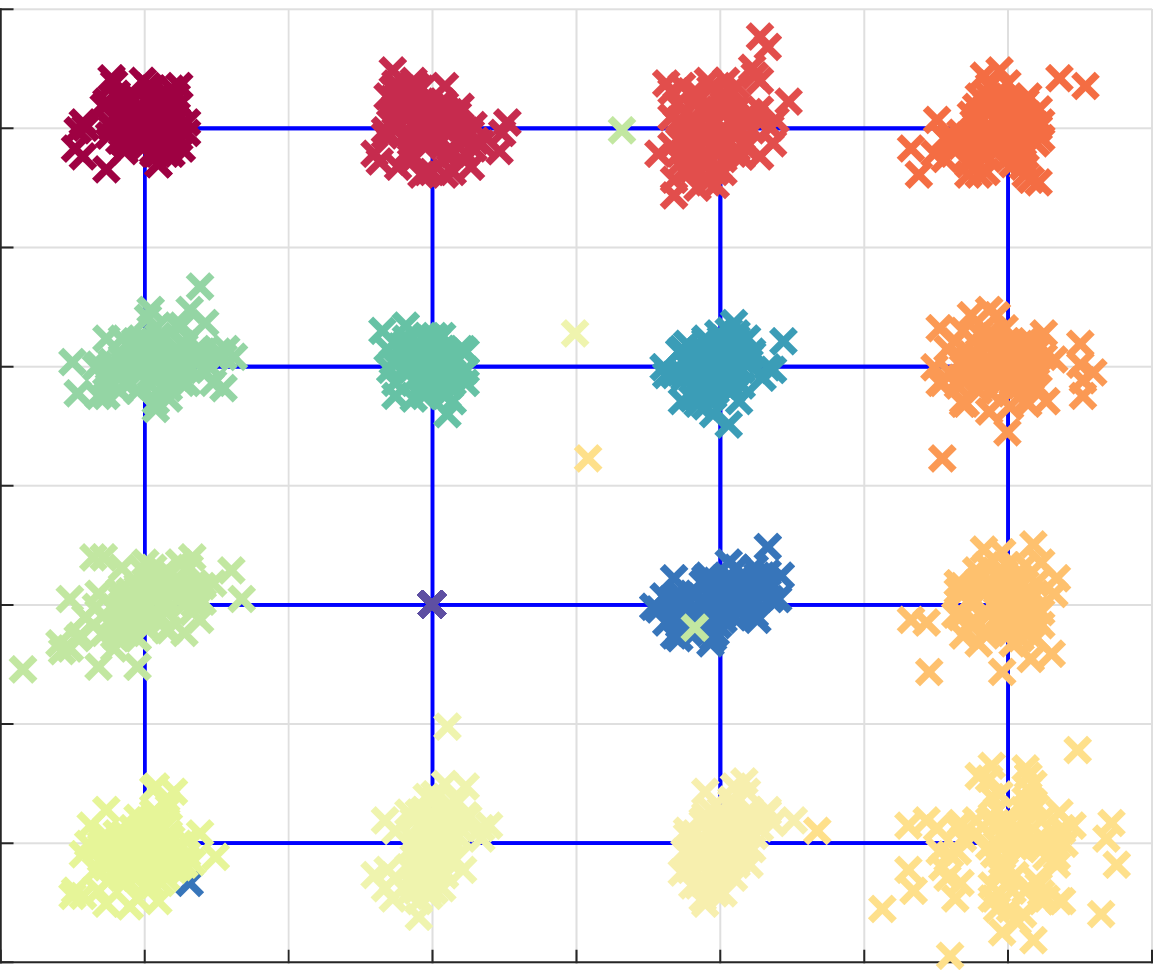}
    \hfill
    \includegraphics[width=0.48\linewidth]{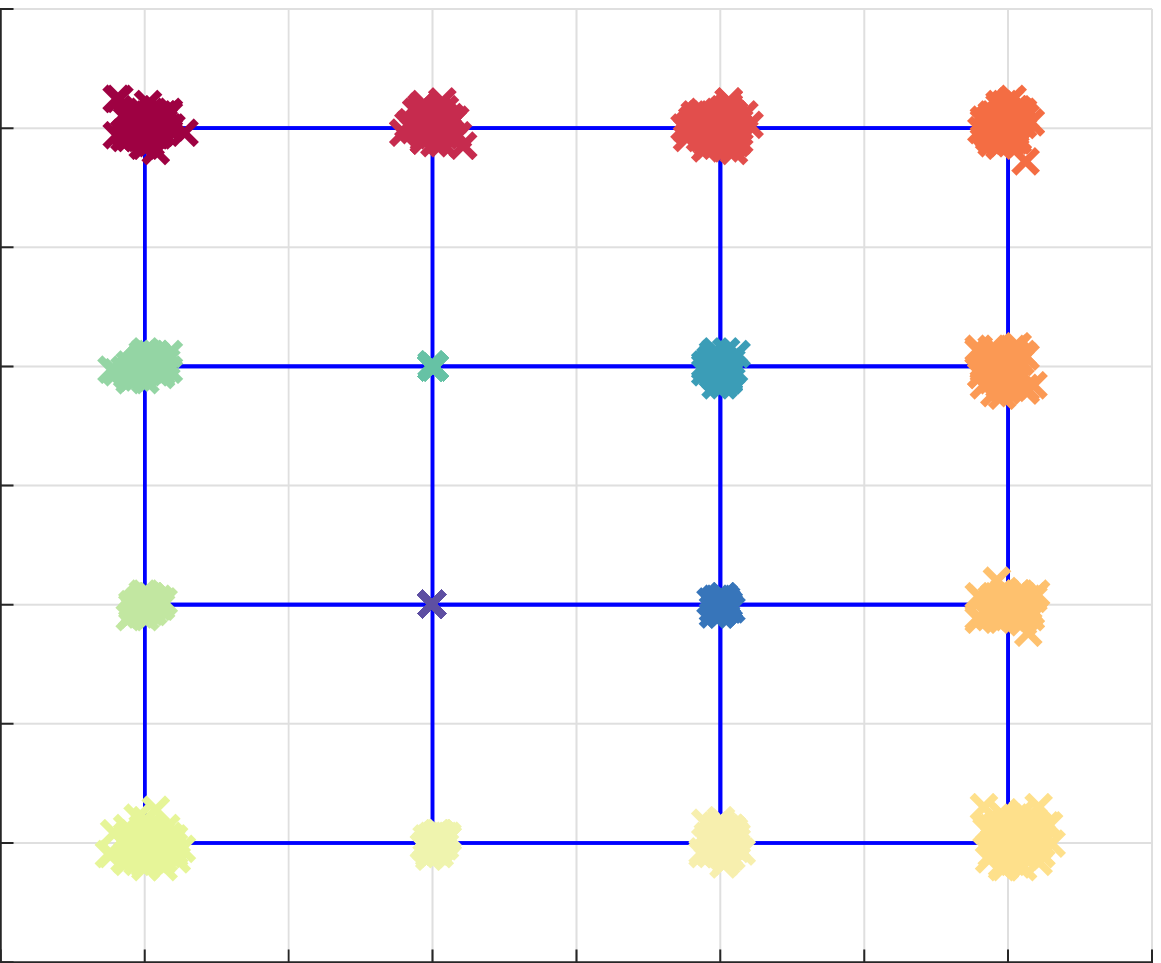}
  \end{minipage}
  \begin{minipage}{\linewidth}
    \includegraphics[width=0.48\linewidth]{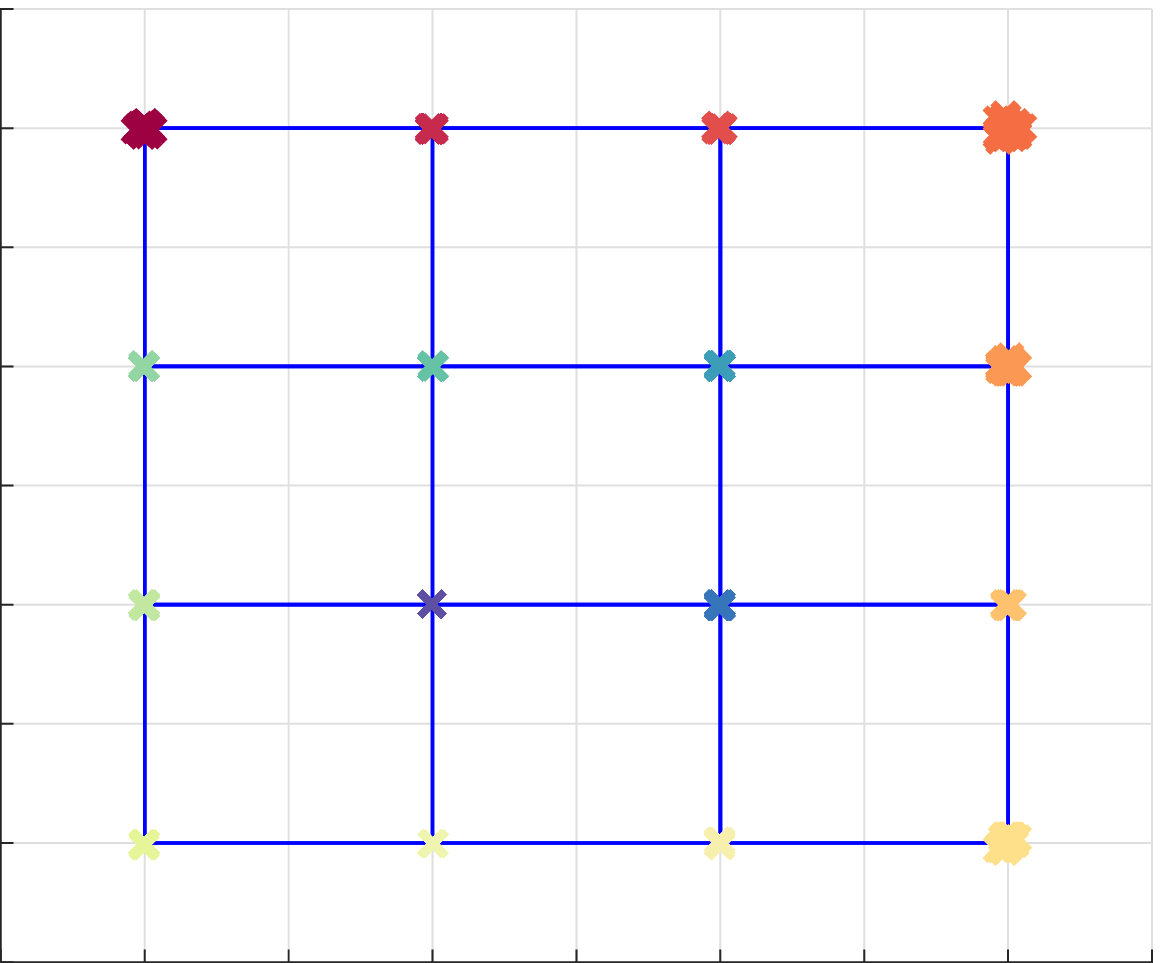} 
    \hfill
    \includegraphics[width=0.48\linewidth]{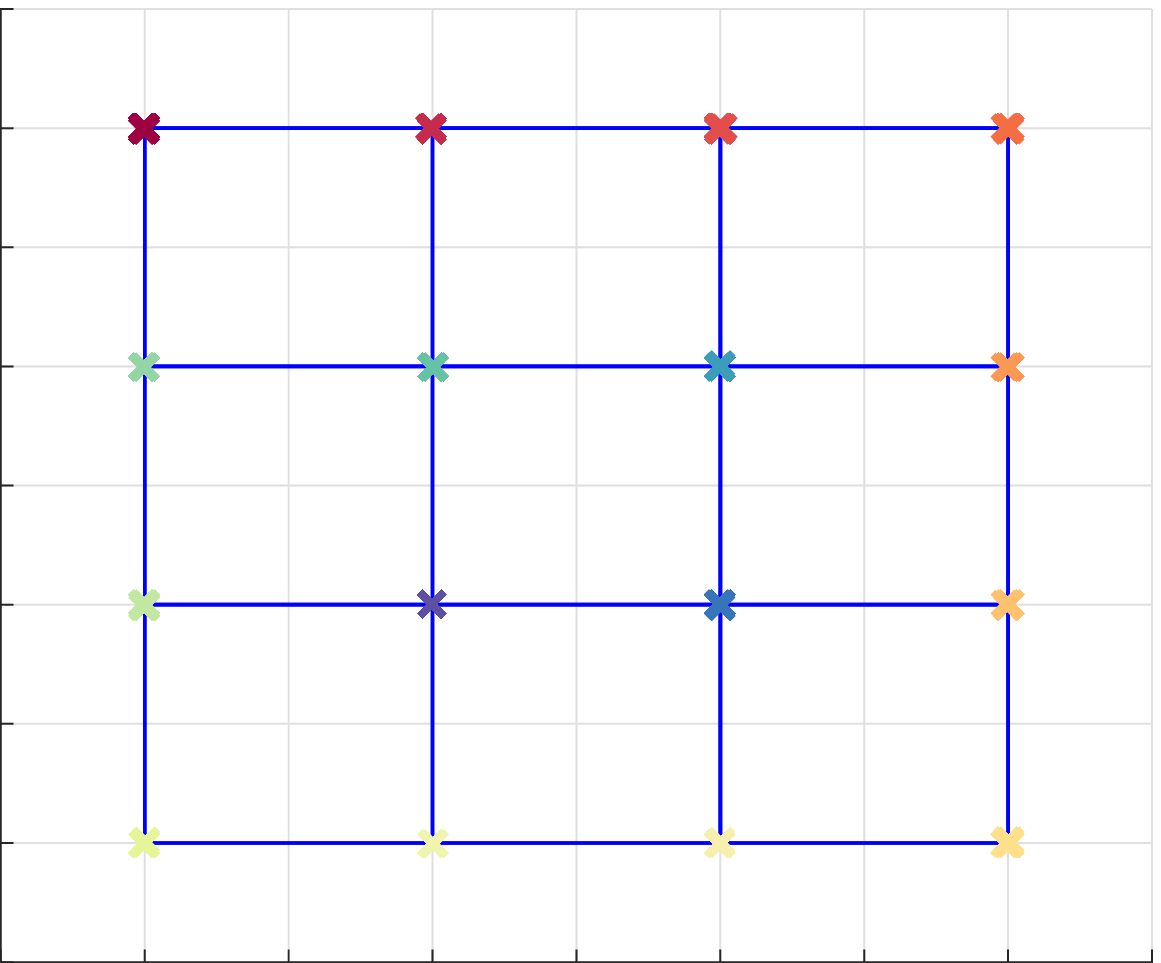}
  \end{minipage}
  \vspace{-4pt}
  }
  \caption[Maximum error]{Node beliefs in LBP iterations: Marginal posterior estimates after iteration 4 (upper-left), 6 (upper-right), 8 (lower-left), and, 10 (lower-right).}
  \label{fig:lbp}
\end{figure}

We use Algorithm~\ref{algnbp} specified in Section~\ref{sec:MonteCarloLBP} for estimating~$\theta$. The MRF model we consider is specified by the pairwise graph $\cal G$ in~\figurename~\ref{fig:scenario} (blue edges). We use $L=100$ points in~\eqref{eqn:locSamples} to represent the local belief densities. We select the sensor data time window length as $t=10$ (starting at time $21$ until $30$ in the scenario in~\figurename~\ref{fig:scenario}). We follow the steps in Algorithm~\ref{algnbp} for $S=16$ iterations. 

A typical run is illustrated in~\figurename~\ref{fig:lbp}. Here, the scatter plot of particles from marginal posteriors are given over iterations.  Note that the network coordinate system is established by sensor~$1$ through its informative prior, and, LBP emanates this information towards the outer nodes while learning the edge potentials using node-wise separable likelihood evaluations $\psi_{i,j}(\theta_i^{(l)},\theta_j^{(l)})$ given in \eqref{eqn:edgepot}~and~\eqref{eqn:quadpotwnorm}.

\begin{figure}[t!]
  \centerline{
    \includegraphics[width=80mm,height=38mm]{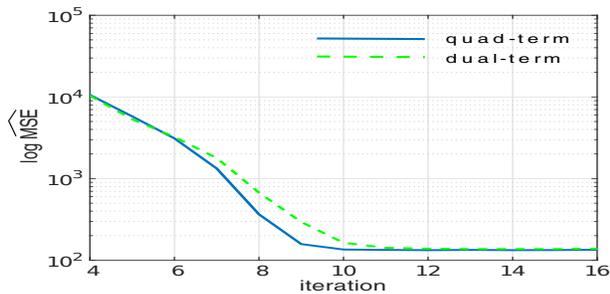}
  }
  \caption[Log error margin]{Log-normalised error margin versus the iteration number $n$.}
  \label{fig:logerrmargin}
\end{figure}

For performance assesment, first, we consider the mean squared error for $\hat \theta$ output by our algorithm, as we have built our discussion on MMSE estimators~in~\eqref{eqn:mmsetheta}. We find this value empirically by taking the average of the squared norm of estimation errors over $100$ Monte Carlo simulations. In \figurename~\ref{fig:logerrmargin}, we present a semi-log plot of this quantity over iterations (blue line). Note that, convergence occurs in less than ten iterations which is a favourable feature. We compare this algorithm with the RFS based dual-term pseudo-likelihood proposed in~\cite{Uney2016}. When evaluating this term, we use a Poisson multi-object model output by using the Gaussian mixture probability hypothesis density (GM-PHD) filter~\cite{Vo2006} with the LGSS model. The averaged MSE performance for the case that dual-term likelihoods are used as edge potentials is depicted with the green dashed line in \figurename~\ref{fig:logerrmargin}. The quad-term approximation is seen to provide  faster convergence with both pseudo-likelihoods leading to an on par accuracy in the steady regime, in this example. The edge update time for the quad-term update averages to $0.601$ per edge per particle compared to $1.312$ for the dual-term update demonstrating its relative efficiency.

\begin{figure}[b!]
  \centerline{
    \includegraphics[width=80mm,height=34mm]{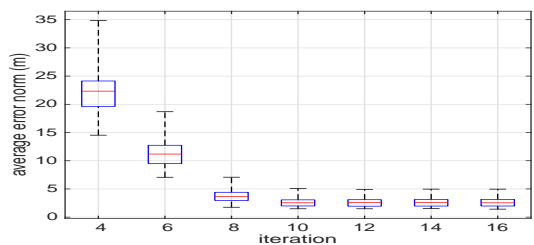}
  }
  \caption[Maximum error]{Localisation miss-distance averaged over nodes versus the iteration number.  $100$ Monte Carlo runs displayed with the boxes centered at the median (red). Edges (blue) indicate the $25$th and $75$th percentiles.}
  \label{fig:averr}
\end{figure}

Note that, the MSE is a network-wide term and the local error norms are smaller. The localisation miss-distance averaged over sensors is given in~\figurename~\ref{fig:averr}. The average error ($\pm$ one standard deviation) in the final step is $2.60 \pm 0.70 m$ with a maximum value of $4.96$ which is less than $0.5 \%$ of the edge distances of $1000m$. These results demonstrate that the proposed scheme is capable of providing self-localisation with favourable accuracy and small error margins. 

\section{Conclusions}
\label{sec:Discussion}
In this work, we have addressed the prohibitive complexity of latent parameter estimation in state space models  when there are many sensors collecting measurements. We proposed a pseudo-likelihood, namely the quad-term node-wise separable likelihood, as an accurate surrogate to the actual likelihood which is extremely costly to evaluate. The separable structure of this quad-term approximation makes it possible to evaluate it using local filtering operations, hence, scale with the number of sensors. 

In order to use the proposed approximation in the case of multiple objects, we employed a parameterised multi-object state space model in which different configurations of the parameter specify different hypothesis of object-to-measurement and object-to-object associations. Specifically, we introduced an empirical Bayesian perspective for evaluating separable likelihoods in this model, using only local Bayesian filtering. This approach substitutes the estimates of the hypothesis variables when evaluating the likelihood of the latent parameter, and, decouples these two inference tasks if the hypothesis variables can be estimated locally. Therefore, it can be extended to general hypothesis variables that capture, for example, variable number of objects $M_k$, less than one probability of detection, i.e., $P_D<1$, and, measurements with false alarm, by 
adapting the assignment problems involved accordingly (see, for example ~\cite{Poore2006}).

The associated posterior distribution is a MRF over which distributed inference is possible using message passing algorithms such as LBP. We specified a particle message passing algorithm for sampling from latent parameter marginals for a linear Gaussian state space model with multiple objects. This algorithm is demonstrated in simulations for sensor self-localisation using point measurements from non-cooperative objects in a potentially GPS denying environment. It is possible to estimate unkown orientation angles as well, by appropriately defining the transform in~\eqref{eqn:coortrans_tau} and taking into account the rotations implied by $\theta_i$s in the LBP steps of Algorithm~\ref{algnbp} (additional details can be found in~\cite{Uney2018}).
\appendix
\subsection{Empirical Bayes parameter likelihood}
\label{sec:empiricalBayesParameterLikelihood}
The empirical Bayes parameter posterior follows from the decomposition of the posterior in~\eqref{eqn:posthetaglobal} using the chain rule of probabilities as
\begin{eqnarray}
 && p(\theta|\vect{Z}^1_{1:t},\ldots,\vect{Z}^N_{1:t})= \label{eqn:posthetaglobal2} \\
 && \sum_{\tau^1_{1:t}}\cdots\sum_{\tau^N_{1:t}} 
p(\theta|\vect{Z}^1_{1:t},\ldots,\vect{Z}^N_{1:t},\tau^{1:N}_{1:t}) p(\tau^{1:N}_{1:t}|\vect{Z}^1_{1:t},\ldots,\vect{Z}^N_{1:t}). \nonumber
\end{eqnarray}

The first term inside the summations is the parameter posterior conditioned on the association variables and the second term is similar to a prior with respect to the first term. The fact that this term is  conditioned on the measurements makes an empirical selection possible as discussed in Section~\ref{sec:statement}. Let us use a similar empirical prior selection approach as used in~\eqref{eqn:empiricalprior}, i.e., 
\begin{eqnarray}
 p(\tau^{1:N}_{1:t}|\vect{Z}^1_{1:t},\ldots,\vect{Z}^N_{1:t}) &= &\prod_{k=1}^t p(\tau^{1:N}_{k}|\vect{Z}^1_{k},\ldots,\vect{Z}^N_{k}) \nonumber \\
 p(\tau^{1:N}_{k}|\vect{Z}^1_{k},\ldots,\vect{Z}^N_{k})&\leftarrow&\delta_{{\bm {\thickbar{\tau}}}^{1:N}_{k-1}}(\tau^{1:N}_{k-1}).
 \label{eqn:empiricalPrior2}
\end{eqnarray}
After substituting from~\eqref{eqn:empiricalPrior2} in~\eqref{eqn:posthetaglobal2}, the parameter posterior is found as

\begin{eqnarray}
&&\mspace{-50mu} p(\theta|\vect{Z}^1_{1:t},\ldots,\vect{Z}^N_{1:t}) \nonumber \\
&=&  p(\theta|\vect{Z}^1_{1:t},\ldots,\vect{Z}^N_{1:t},\tau^{1:N}_{1:t} = {\bm {\thickbar{\tau}}}^{1:N}_{1:t}) \nonumber \\
&\propto& l(\vect{Z}^1_{1:t},\ldots,\vect{Z}^N_{1:t}|\theta, \tau^{1:N}_{1:t}={\bm {\thickbar{\tau}}}^{1:N}_{1:t}) p(\theta) \nonumber
\end{eqnarray}
where the likelihood in the last line is given by~\eqref{eqn:conditionalcentlhood}~and~\eqref{eqn:factcent_emp} in Section~\ref{sec:statement}.
\vspace{-4pt}
\subsection{Proof of Proposition~\ref{prop:quadupperbound}}
\label{sec:proofProp1}
\begin{proof}
Let us expand the KLD term in~\eqref{eqn:quadupperbound} by substituting its arguments given in \eqref{eqn:factcent}, \eqref{eqn:quadapprox}~and~\eqref{eqn:quadscale}:
\begin{eqnarray}
&& \mspace{-30mu} D\left(p({\vect Z}^i_k,{\vect Z}^j_k|{\vect Z}^i_{1:k-1},{\vect Z}^j_{1:k-1},\theta)||q({\vect Z}^i_k, {\vect Z}^j_k| {\vect Z}^i_{1:k-1}, {\vect Z}^j_{1:k-1},\theta) \right) \notag \\
&=&\int \mathrm{d}{\vect Z}^i_{1:k}\mathrm{d}{\vect Z}^j_{1:k} \mathrm{d}\theta\,p({\vect Z}^i_{1:k}, {\vect Z}^j_{1:k}, \theta ) \notag \\
&& \mspace{100mu} \times \log \frac{p({\vect Z}^i_k, {\vect Z}^j_k| {\vect Z}^i_{1:k-1}, {\vect Z}^j_{1:k-1},\theta)}{q({\vect Z}^i_k, {\vect Z}^j_k| {\vect Z}^i_{1:k-1}, {\vect Z}^j_{1:k-1},\theta)}\notag \\
&=&\int \mathrm{d}{\vect Z}^i_{1:k}\mathrm{d}{\vect Z}^j_{1:k}\mathrm{d}\theta\, p({\vect Z}^i_{1:k}, {\vect Z}^j_{1:k},\theta ) \notag \\
&&\mspace{50mu} \times \frac{1}{2}  \left(
\log  \frac{ p({\vect Z}^i_k, {\vect Z}^j_k,{\vect Z}^i_{1:k-1} | {\vect Z}^j_{1:k-1},\theta )}{ p({\vect Z}^i_k, {\vect Z}^j_k|{\vect Z}^j_{1:k-1},\theta )p({\vect Z}^i_{1:k-1} |{\vect Z}^j_{1:k-1},\theta)}   \right.
\notag \\
&&\mspace{75mu} \left. + \log \frac{p({\vect Z}^i_k, {\vect Z}^j_k,{\vect Z}^j_{1:k-1} | {\vect Z}^i_{1:k-1},\theta )}{ p({\vect Z}^i_k, {\vect Z}^j_k|{\vect Z}^i_{1:k-1},\theta )p({\vect Z}^j_{1:k-1}|{\vect Z}^i_{1:k-1},\theta)} \right. \notag \\
&&\mspace{100mu} \left. + 2 \log \kappa_k(\theta)\right. \Big)
\label{eqn:kldquadsumolog} \\
&=&\frac{1}{2}\left( I({\cal Z}_k^j,{\cal Z}^i_k; {\cal Z}^i_{1:k-1}|{\cal Z}^j_{1:k-1},\Theta) \right. \notag \\ 
&&\mspace{25mu} \left. +I({\cal Z}^j_k,{\cal Z}^i_k;{\cal Z}^j_{1:k-1}|{\cal Z}^i_{1:k-1},\Theta) \right) + E\{\log \kappa_k(\theta) \}.
\label{eqn:kldquad1}
\end{eqnarray}
In the equations above, $\kappa_k(\theta)$ is a normalisation constant given by~\eqref{eqn:quadscale}. 
Eq.\eqref{eqn:kldquadsumolog} is obtained after multiplying both the numerator and the denominator of the quotient inside the logarithm by $p({\vect Z}^i_{1:k-1}|{\vect Z}^j_{1:k-1},\theta)p({\vect Z}^j_{1:k-1}|{\vect Z}^i_{1:k-1},\theta)$ and a rearrangement of the terms. The definition of MI~\cite{Cover1991} results with the first two terms in Eq.\eqref{eqn:kldquad1}. The last term is the expectation of the normalisation constant over the joint distribution of the sensor histories ${\cal Z}^i_{1:k-1}$~and~${\cal Z}^j_{1:k-1}$, and,~$\Theta$.

Let us now consider the normalisation constant:
\begin{eqnarray}
 \kappa_k(\theta) &=& \int \mathrm{d}{\vect Z}^i_k \mathrm{d}{\vect Z}^j_k \left(  p({\vect Z}^i_k , {\vect Z}^j_k|{\vect Z}^j_{1:k-1},\theta )p({\vect Z}^i_k, {\vect Z}^j_k| {\vect Z}^i_{1:k-1},\theta ) \right)^{1/2} \notag \\
 &\leq&  \left( \int\mathrm{d}{\vect Z}^i_k \mathrm{d}{\vect Z}^j_k  p({\vect Z}^i_k , {\vect Z}^j_k|{\vect Z}^j_{1:k-1},\theta ) \right)^{1/2} \notag \\
 &&\mspace{40mu}\times \left( \int\mathrm{d}{\vect Z}^i_k \mathrm{d}{\vect Z}^j_k  p({\vect Z}^i_k , {\vect Z}^j_k|{\vect Z}^i_{1:k-1},\theta )\right)^{1/2}
 \label{eqn:holderinq2}\\
 &=&1 \notag.
\end{eqnarray}
The inequality \eqref{eqn:holderinq2} follows from H\"older's Inequality. Consequently, the last term in \eqref{eqn:kldquad1} is non-positive, and,~\eqref{eqn:quadupperbound} is obtained.
\end{proof}
\vspace{-4pt}
\subsection{Proof of Corollary~\ref{cor:quadupperbound}}
\label{sec:proofCorollary1}
\begin{proof}
We apply the chain rule of information to the MI terms on the RHS of \eqref{eqn:quadupperbound} leading to

\begin{multline}
 I({\cal Z}_k^i, {\cal Z}_k^j; {\cal Z}^i_{1:k-1} | {\cal Z}^j_{1:k-1}, \Theta) = \\
 I({\cal Z}_k^j;  {\cal Z}^i_{1:k-1} | {\cal Z}^j_{1:k-1}, \Theta)
  + I({\cal Z}^i_k;  {\cal Z}^i_{1:k-1} | {\cal Z}^j_{1:k}, \Theta), \label{eqn:quadboundmi1}
 \end{multline}
 and,
 \begin{multline}
 I({\cal Z}^i_k,{\cal Z}^j_k; {\cal Z}^j_{1:k-1} | {\cal Z}^i_{1:k-1}, \Theta ) = \\
 I({\cal Z}^i_k;  {\cal Z}^j_{1:k-1} | {\cal Z}^i_{1:k-1}, \Theta) 
 + I(Z^j_k;  {\cal Z}^j_{1:k-1} | Z^i_{1:k}, \Theta). \label{eqn:quadboundmi2}
\end{multline}
The MI terms on the RHSs of the equations above are for random variables which form Markov chains with the current the state variable ${\cal X}_k$. Consider the (conditional) chains ${\cal Z}^j_k \leftrightarrow {\cal X}_k \leftrightarrow {\cal Z}^i_{1:k-1} |  {\cal Z}^j_{1:k-1},\Theta $ and ${\cal Z}^i_k \leftrightarrow {\cal X}_k \leftrightarrow {\cal Z}^i_{1:k-1}|  {\cal Z}^j_{1:k-1},\Theta$ for the RHS of Eq.\eqref{eqn:quadboundmi1}. The Data Processing Inequality~\cite{Cover1991} applied to these terms lead to 
\begin{eqnarray}
 && \mspace{-30mu}I({\cal Z}^i_k,{\cal Z}^j_k; {\cal Z}^i_{1:k-1} | {\cal Z}^j_{1:k-1},\Theta ) \notag \\ &\leq& I({\cal X}_k; {\cal Z}^i_{1:k-1} | {\cal Z}^j_{1:k-1},\Theta ) + I({\cal X}_k^i;  {\cal Z}^i_{1:k-1} |{\cal Z}^j_k, {\cal Z}^j_{1:k-1},\Theta) \notag \\
 \mspace{20mu}&=& H({\cal X}_k|{\cal Z}^j_{1:k-1},\Theta ) - H({\cal X}_k| {\cal Z}^i_{1:k-1}, {\cal Z}^j_{1:k-1},\Theta )  \label{eqn:quadboundmi3} \\
 &&\mspace{-20mu} + H({\cal X}_k|{\cal Z}^j_k, {\cal Z}^j_{1:k-1},\Theta ) - H({\cal X}_k|{\cal Z}^j_k, {\cal Z}^j_{1:k-1},{\cal Z}^i_{1:k-1},\Theta) \notag
\end{eqnarray}
A similar break down of Eq.\eqref{eqn:quadboundmi2} results with
\begin{eqnarray}
 && \mspace{-30mu} I({\cal Z}^i_k,{\cal Z}^j_k; {\cal Z}^j_{1:k-1} | {\cal Z}^i_{1:k-1},\Theta )\notag \\ 
 &&\mspace{0mu} \leq H({\cal X}_k| {\cal Z}^i_{1:k-1},\Theta) - H({\cal X}_k| {\cal Z}^i_{1:k-1}, {\cal Z}^j_{1:k-1},\Theta) \label{eqn:quadboundmi4} \\
  &&\mspace{-20mu}+ H({\cal X}_k|  {\cal Z}^i_k, {\cal Z}^i_{1:k-1},\Theta ) - H({\cal X}_k|  {\cal Z}^i_k, {\cal Z}^i_{1:k-1}, {\cal Z}^j_{1:k-1},\Theta ). \notag
\end{eqnarray}
Substituting from \eqref{eqn:quadboundmi3} and \eqref{eqn:quadboundmi4} into \eqref{eqn:quadupperbound} results with \eqref{eqn:corolH}.
\end{proof}
\subsection{Comparison of the quad-term and dual-term updates}
\label{sec:kldcomp}
Let us compare the KLDs of the quad-term and dual-term updates. The KLD of the dual-term in~\eqref{eqn:dualtermupdate} is given by~\cite{Uney2016}
 \begin{multline}
 \mspace{-20mu}
D(p(\vect{Z}_k^i, \vect{Z}_k^j| \vect{Z}_{1:k-1}^i,\vect{Z}_{1:k-1}^j,\theta)||u(\vect{Z}_k^i, \vect{Z}_k^j| \vect{Z}_{1:k-1}^i,\vect{Z}_{1:k-1}^j,\theta) )  \\ 
= I({\cal Z}_k^j; {\cal Z}_{1:k-1}^j | {\cal Z}_{1:k-1}^i,{\Theta} ) + I({\cal Z}_k^i;{\cal Z}_{1:k-1}^i| {\cal Z}_{1:k-1}^j  ,{\Theta} ) \\ + I({\cal Z}_k^i;{\cal Z}_k^j|{\cal Z}_{1:k-1}^i, {\cal Z}_{1:k-1}^j ,{\Theta} ).
 \label{eqn:dualkld}
\end{multline}
The MI terms on the RHS of the above equation can be expanded using that $I({\cal A};{\cal B}|{\cal C})=H({\cal A}|{\cal C})-H({\cal A}|{\cal C},{\cal B})$~\cite{Cover1991}. The third term can be expanded in alternative ways as follows:
\begin{eqnarray}
 &&\mspace{-40mu} I({\cal Z}_k^i;{\cal Z}_k^j|{\cal Z}_{1:k-1}^i, {\cal Z}_{1:k-1}^j ,{\Theta} ) \nonumber \\
&\mspace{-50mu} \overset{\text{(a)}}{=}& \mspace{-20mu}H({\cal Z}^i_k|{\cal Z}_{1:k-1}^i, {\cal Z}_{1:k-1}^j ,{\Theta}) - H({\cal Z}^i_k|{\cal Z}_{1:k-1}^i, {\cal Z}_{1:k}^j ,{\Theta}) 
 \label{eqn:lastMIterma} \\
 &\mspace{-50mu} \overset{\text{(b)}}{=}& \mspace{-20mu} H({\cal Z}^j_k|{\cal Z}_{1:k-1}^i, {\cal Z}_{1:k-1}^j ,{\Theta}) - H({\cal Z}^j_k|{\cal Z}_{1:k-1}^i, {\cal Z}_{1:k}^i ,{\Theta}) 
  \label{eqn:lastMItermb}
\end{eqnarray}
After decomposing the first two terms on the RHS of~\eqref{eqn:dualkld} similarly and adding to the average of \eqref{eqn:lastMIterma}~and~\eqref{eqn:lastMItermb} (which equals to the third term), we obtain
 \begin{eqnarray}
 D(p||u) &=& \Delta u^+ -\Delta u^- \label{eqn:dualkld_exp} \\
  \Delta u^+ & \triangleq &  H({\cal Z}^j_k|{\cal Z}^i_{1:k-1},{\Theta}) + H({\cal Z}^i_k|{\cal Z}^j_{1:k-1},{\Theta})  \nonumber \\
\Delta u^- & \triangleq & -\frac{1}{2}\left[ H({\cal Z}^i_k|{\cal Z}_{1:k-1}^i, {\cal Z}_{1:k-1}^j ,{\Theta}) \right. \nonumber \\
&&\left.\mspace{-50mu} + H({\cal Z}^j_k|{\cal Z}_{1:k-1}^i, {\cal Z}_{1:k-1}^j ,{\Theta}) \right. \nonumber \\
 &&\left.\mspace{-50mu} + H({\cal Z}^i_k|{\cal Z}_{1:k-1}^i, {\cal Z}_{1:k}^j ,{\Theta}) + H({\cal Z}^j_k|{\cal Z}_{1:k}^i, {\cal Z}_{1:k-1}^j ,{\Theta}) \right] \nonumber
\end{eqnarray}

Now, let us consider~\eqref{eqn:quadupperbound}. After substituting~\eqref{eqn:quadboundmi1}~and~\eqref{eqn:quadboundmi2} in the RHS and expanding the MI terms as above, one obtains

\begin{eqnarray}
 D(p||q) &\leq& \Delta q^{+} - \Delta u^- \label{eqn:quadkld_exp} \\
\Delta q^{+} &\triangleq& \frac{1}{2}\left[ H({\cal Z}^j_k|{\cal Z}^j_{1:k-1},{\Theta}) + H({\cal Z}^i_k|{\cal Z}^i_{1:k-1},{\Theta}) \right. \nonumber \\
&& \left. + H({\cal Z}^j_k|{\cal Z}^i_{1:k},{\Theta}) + H({\cal Z}^i_k|{\cal Z}^j_{1:k},{\Theta}) \right] \nonumber
\end{eqnarray}

Now, let us compare~\eqref{eqn:quadkld_exp}~and~\eqref{eqn:dualkld_exp}: The negative weighted terms are equal, so, the difference of the positive weighted terms are considered:
\begin{eqnarray}
 \Delta u^+ - \Delta q^+ &=&  H({\cal Z}^j_k|{\cal Z}^i_{1:k-1},{\Theta}) -H({\cal Z}^j_k|{\cal Z}^j_{1:k-1},{\Theta})
 \label{eqn:deltadiff}\\
 &&+ H({\cal Z}^i_k|{\cal Z}^j_{1:k-1},{\Theta}) - H({\cal Z}^i_k|{\cal Z}^i_{1:k-1},{\Theta})  + \epsilon \nonumber \\
 \epsilon &\triangleq  &H({\cal Z}^j_k|{\cal Z}^i_{1:k-1},{\Theta}) -  H({\cal Z}^j_k|{\cal Z}^i_{1:k},{\Theta}) \nonumber \\
&& + H({\cal Z}^i_k|{\cal Z}^j_{1:k-1},{\Theta}) - H({\cal Z}^i_k|{\cal Z}^j_{1:k},{\Theta})
\nonumber
\end{eqnarray}

Properties of differential entropy~\cite{Cover1991} suggest that $\epsilon \geq 0$ regardless of the problem setting (as conditioning reduces entropy). Under normal sensing conditions it is reasonable to expect that $\epsilon > 0$ holds as sensor measurements are highly informative on all the variables at time $k$. For sensors with similar sensing capabilities, it is also reasonable to expect that their measurement history are interchangeable. In other words,
\begin{multline}
 H({\cal Z}^i_k|{\cal Z}^j_{1:k-1},{\Theta}) + H({\cal Z}^j_k|{\cal Z}^i_{1:k-1},{\Theta}) = \\  H({\cal Z}^j_k|{\cal Z}^j_{1:k-1},{\Theta}) + H({\cal Z}^i_k|{\cal Z}^i_{1:k-1},{\Theta})
 \label{eqn:eqsenscond}
\end{multline}
holds in which case~\eqref{eqn:deltadiff} is nonzero and $D(p||u)>D(p||q)$. This condition can be relaxed for a difference of $\epsilon$ between the RHS and LHS of~\eqref{eqn:eqsenscond}.

Comparison in terms of entropy upper bounds is more straightforward. Let us consider~\eqref{eqn:corolH}~and~Corollary 4.2~in~\cite{Uney2016} which we repeat here for convenience:
\begin{eqnarray}
 D(p||u) &\leq& H_u^+ - H_u^-, \label{eqn:quadkld_Hbound} \\
 H_u^+ &=& H( {\cal X}_k|{\cal Z}^j_{1:k-1} , {\Theta} ) + H({\cal X}_k|{\cal Z}^i_{1:k-1},{\Theta}), \nonumber \\
  H_u^- &=& H({\cal X}_k | {\cal Z}_{1:k-1}^i, {\cal Z}_{1:k-1}^j ,{\Theta}) + \nonumber \\
  && \mspace{-30mu}\max\{ H({\cal X}_k | {\cal Z}_{1:k-1}^i, {\cal Z}_{1:k}^j ,{\Theta}),
   H({\cal X}_k | {\cal Z}_{1:k-1}^j, {\cal Z}_{1:k}^i ,{\Theta}) \}. \nonumber
\end{eqnarray}
For sensors of identical capabilities, both terms in the maximisation should be identical as the accuracy of the state estimate should not differ for using either of the sensors' current measurement in addition to the histories of both. As a result, we can replace $H_u^-$ with 
\begin{multline}
  H_u^- = H({\cal X}_k | {\cal Z}_{1:k-1}^i, {\cal Z}_{1:k-1}^j ,{\Theta}) + \nonumber \\
 \frac{1}{2} \left( H({\cal X}_k | {\cal Z}_{1:k-1}^i, {\cal Z}_{1:k}^j ,{\Theta}) +   H({\cal X}_k | {\cal Z}_{1:k-1}^j, {\cal Z}_{1:k}^i ,{\Theta}) \right). \nonumber
\end{multline}
Now, note that the sum of the positive weighted terms in the RHS of~\eqref{eqn:corolH} (let us denote by $H_q^+$) is smaller than $H_u^+$ owing to that conditioning reduces entropy~\cite{Cover1991}. The sum of the negative weighted terms (let us denote by $H_q^-$), for the case ,equals to $H_u^-$, leading to 
\begin{equation}
 H_u^+ - H_u^- - (H_q^+ - H_q^-) > 0, \nonumber
\end{equation}
which indicates that the entropy bound of the quad-term update is smaller than that for the dual-term update.
\vspace{-4pt}
\subsection{Local ML estimate of object correspondances}
\label{sec:mlgamma}
The semantic of the object correspondance $\gamma_k$ is slightly different from that of the random permutation variables used in identity management~\cite{Guibas2008} in a way closer to data association because local (track) identities are synonymous with the measurements they are associated with instead of signal features etc. Therefore, in our model, $\gamma_k$ is uniquely defined when given $ \tau^i_k$~and~$\tau^j_k$, and, $\gamma_{1},\ldots,\gamma_{k}$ are mutually independent. The ML estimate of $\gamma_k$ using the likelihood for data set $({\vect{Z}}^i_k,{\vect Z}^j_{1:k})$ is also conditioned on the association configuration $(\hat  \tau^i_k, \hat  \tau^i_k )$ and $\theta$ is given by
\begin{equation}
 \hat \gamma_k =\arg \max_{\gamma_k \in {\cal S}_M} \log l({\vect Z}^i_k,{\vect Z}^j_{1:k} |\gamma_k, \tau^i_k=\hat \tau^i_k, \tau^j_k= \hat \tau^j_k, \theta ) \label{eqn:identityproblem}
 \end{equation}
 where ${\cal S}_M$ is the set of $M$-permutations. This likelihood can be decomposed using the chain rule of probability as
 \begin{eqnarray}
  && \mspace{-20mu}l({\vect Z}^i_k,{\vect Z}^j_{1:k} |\gamma_k, \tau^i_k, \tau^j_k, \theta ) \nonumber \\
  &= &p({\vect Z}^i_k,{\vect Z}^j_{k} |{\vect Z}^j_{1:k-1}, \gamma_k, \tau^i_k, \tau^j_k, \theta )p({\vect Z}^j_{1:k-1} |\gamma_k, \tau^i_k, \tau^j_k,\theta), \nonumber \\
  &=& p({\vect Z}^i_k,{\vect Z}^j_{k} |{\vect Z}^j_{1:k-1}, \gamma_k, \tau^i_k, \tau^j_k, \theta )p({\vect Z}^j_{1:k-1} |\theta),
 \end{eqnarray}
where the second equality follows from the independence of sensor~$j$'s measurements up to time $k-1$ from the association variables at time $k$. The first term on the RHS is easily identified as~\eqref{eqn:rij_o} evaluated for $\gamma_k$ when $m'= \gamma_k( \tau^i_k(o))$. This term, when subsituted in~\eqref{eqn:identityproblem}, leads to a $2$-D assignment problem given by
\begin{eqnarray}
  \hat \gamma_k&=& \arg \max_{  {\gamma_k}\in {\cal S}_M } \sum_{o=1}^M  d( o ,m'= \gamma_k(\hat \tau^i_k(o)) ) \nonumber \\
 d(o,m') &\triangleq & \log r^k_{ij,o}( z_{k,o}^i ,\theta), \nonumber
\end{eqnarray}
for $o,m'=1,\ldots,M$, where $r_{ij,o}^k$ is given in~\eqref{eqn:rij_o}.

After finding the $M^2$ costs above, this problem can be solved using the auction algorithm~\cite{Bertsekas1988} in polynomial time with $M$. This algorithm finds the $M$ pairs corresponding to the best permutation $\hat \gamma_{k}$. We use a similar approach for the data association problem in Bayesian filtering in Section~\ref{sec:ssfilter}.
\ifCLASSOPTIONcaptionsoff
 \newpage
\fi
\section*{Acknowledgment}
\addcontentsline{toc}{section}{Acknowledgment}
The authors would like to thank Prof. Arnaud Doucet for discussions on pseudo-likelihood methods in state space models, and, Dr.~Sinan~Y\.{i}ld\.{i}r\.{i}m for discussions on the general multi-object tracking model in~\cite{Jiang2015}~and~\cite{Yildirim2015a}8.

\vfill
\end{document}